\documentclass[11pt]{article}
\usepackage{a4wide}                      
\usepackage{amsfonts}                    
\usepackage{amsmath2000}                 
\usepackage{amssymb}                     
\usepackage{latexsym}                    
\usepackage{epsf}                        

\newcommand{\sect}[1]{\setcounter{equation}{0}\section{#1}}
\renewcommand{\theequation}{\arabic{section}.\arabic{equation}}

\def\be{\begin{equation}}
\def\ee{\end{equation}}
\def\bea{\begin{eqnarray}}
\def\eea{\end{eqnarray}}
\def\nn{\nonumber \\}
\def\vsp#1{\vspace{#1}}
\def\hsp#1{\hspace{#1}}

\def\part{\partial}
\def\tfrac#1#2{{\textstyle{\frac{#1}{#2}}}}
\def\half{\tfrac{1}{2}}
\def\x{\times}

\def\incl{\mbox{i}}
\def\STr{\mbox{STr}}
\def\aprime{{\alpha^\prime}}
\def\sqaprime{{\sqrt{\alpha^\prime}}}
\def\R{\ensuremath{\mathbb{R}}}
\def\Z{\ensuremath{\mathbb{Z}}}

\def\tB{\tilde B} 
\def\tF{\tilde F}

\def\dA{\dot A}
\def\dX{\dot X}
\def\dZ{\dot Z}

\def\cD{{\cal D}}  

\def\Ortin{Ort{\'\i}n}

\def\mn{{\mu\nu}}

\def\a9{{a9}}

\def\makeatletter{\catcode`\@=11}
\makeatletter
\def\mathbox#1{\hbox{$\m@th#1$}}%
\def\math@ccstyles#1#2#3#4#5#6#7{{\leavevmode
      \setbox0\mathbox{#6#7}%
      \setbox2\mathbox{#4#5}%
      \dimen@ #3%
      \baselineskip\z@\lineskiplimit#1\lineskip\z@
      \vbox{\ialign{##\crcr
             \hfil \kern #2\box2 \hfil\crcr
             \noalign{\kern\dimen@}%
             \hfil\box0\hfil\crcr}}}}
\def\mathaccstyles{\math@ccstyles\maxdimen}
\def\maththroughstyles{\math@ccstyles{-\maxdimen}}
\def\unity%
 {\maththroughstyles{.45\ht0}\z@\displaystyle {\mathchar"006C}\displaystyle 1}
\begin{document}

\rightline{\small DCPT-02/31} 
\rightline{\small FFUOV-02/05}
\rightline{1 August 2002}
\vspace{1.3truecm}

\centerline{\Large \bf On the Dielectric Effect for Gravitational Waves }
\vspace{1truecm}

\centerline{
    {\bf Bert Janssen${}^{a,}$}\footnote{E-mail address: 
                                  {\tt bert.janssen@durham.ac.uk} }
    {\bf and} 
    {\bf Yolanda Lozano${}^{b,}$}\footnote{E-mail address: 
                                  {\tt yolanda@string1.ciencias.uniovi.es}}
                                                            }
\vspace{.4cm}
\begin{center}
{\it ${}^a$Department of Mathematical Sciences,  \\
           South Road, Durham DH1 3LE, United Kingdom} 
\end{center}

\begin{center}
{\it ${}^b$Departamento de F{\'\i}sica,  Universidad de Oviedo \\  
          Avda.~Calvo Sotelo 18, 33007 Oviedo, Spain}
\end{center}
\vspace{2truecm}

\centerline{\bf ABSTRACT}
\vspace{.5truecm}

\noindent
We argue that the dielectric effect, mostly studied for systems of coincident
D-branes, is also extendible to configurations of multiple gravitational 
waves. 
We provide some evidence that Matrix string theory has an alternative
interpretation as describing also, in the static gauge, multiple Type
IIA gravitational waves. Starting with the linearised action of Matrix 
string theory in a weakly curved background, we identify the non-Abelian 
couplings of multiple coinciding gravitons to 
weak background fields, both in Type IIA and in Type IIB, and
we use them to study various dielectric configurations from the point of
view of the expanding gravitons. We also identify, in the Abelian 
limit, some non-perturbative correction terms to the Abelian gravitational 
wave action.

\newpage
\sect{Introduction}

The dielectric effect, i.e.~the idea that a collection of branes
can undergo an expansion into a higher-dimensional brane under the
influence of an external flux tube, has recently attracted a lot of attention. 
This effect was first explored in \cite{emparan1}, where Emparan constructed 
solutions of the Abelian Born-Infeld theory, describing a single D$p$-brane 
with $N$ units of Born-Infeld flux dissolved in its world volume, 
that could be interpreted in terms of expanded fundamental strings or
expanded 
D$(p-2)$-branes. These 
solutions carry no global charge with respect to the background 
R-R $(p+1)$-form potential that needs to be switched on in order to prevent
the D$p$-brane from collapsing, 
but do have a non-zero local charge, which results in a non-zero 
dipole moment, that renders the configuration stable.

A microscopic description of this effect, from the point of
view of the lower-dimensional D$(p-2)$-branes, was provided some years
later by Myers \cite{myers}. He observed that a system of coincident
D-branes can develop multipole moment couplings to R-R fields that would
normally be associated to higher-dimensional branes. From this point of view,
the expansion is due to the fact that the transverse coordinates to
the coincident D-branes are matrix-valued, which allows the D-branes to couple 
to background R-R fields $C^{(q-1)}$, with $q>p$ ($p,q$ even (odd) in
IIA (IIB)) \cite{myers,TVR}. Myers constructed a solution of the non-Abelian 
world volume effective action, describing $N$ coinciding D$(p-2)$-branes under 
the influence of an external R-R $(p+2)$-form field strength. He interpreted 
this non-Abelian solution as the expansion of the $N$ D$(p-2)$ branes into a 
D$p$-brane of topology $\R^{1,p-2}\times S^2_{NC}$, where $S^2_{NC}$ denotes the
non-commutative, or fuzzy 2-sphere. The D$p$-brane is, as in the Abelian 
description, uncharged with respect to the R-R $(p+1)$-form potential, but has a
non-zero dipole moment. Hence the name of dielectric D$p$-brane, introduced by 
Myers. Other expanded brane configurations, involving more general fuzzy cosets,
have been considered in \cite{TV}. 
 
Both descriptions of the dielectric effect, Abelian and non-Abelian, have 
complementary ranges of validity \cite{myers}, that coincide in the limit where 
the number $N$ of D$(p-2)$-branes goes to infinity, in which the non-commutative
nature of the fuzzy two-sphere is lost. 

Supergravity solutions corresponding to expanding branes have been 
constructed, specifically for D4-branes \cite{CHC, emparan2} and 
fundamental strings \cite{emparan2} expanding into a dielectric D6-brane 
of topology $\R^{1,4} \x S^2$ and  $\R^{1,1} \x S^5$ respectively. Again
here, the physical properties of these supergravity solutions, such as 
their stability and the size of the fuzzy spheres, match with the Abelian 
world volume analysis of the dielectric D6-brane \cite{CHC, BS}. 

A microscopical description of dielectric configurations involving 
fundamental strings has also been proposed. Expanding fundamental 
strings have been studied at the non-Abelian, microscopic level in 
\cite{schiappa, BJL} (see also \cite{pedro}), using Matrix string theory in a 
weakly curved background. Matrix string theory \cite{DVV} turns out to give the 
right description of fundamental strings in terms of matrix-valued coordinates,
and the linear couplings to weak background fields generate precisely the 
non-Abelian couplings responsible for the dielectric effect. Dielectric solutions
of fundamental strings expanding into D3-branes and D4-branes 
were constructed in \cite{BJL}.

We also argued in \cite{BJL}, and we will provide further evidence in this
paper, that Matrix string theory, in the static gauge, describes coinciding
Type IIA gravitational waves. Indeed, we will see that Matrix string 
theory has an alternative interpretation as describing also the dynamics of 
multiple gravitational waves.  Since Matrix string theory describes string 
states with fixed light cone momentum, it is not surprising that, in some
limit, it can effectively describe gravitational waves.

A possible dielectric effect for gravitational waves seems to be related 
to another topic that has recently received a lot of attention: giant gravitons
\cite{GST}. One can roughly think of a giant graviton configuration in terms 
of gravitons expanded into a $p$-brane with topology $\R\times S^p$, carrying 
angular momentum and dipole moment.
This configuration breaks supersymmetry in the same way as a point-like graviton
propagating with the same velocity, but in $AdS_m \times S^{p+2}$ spacetimes,
where the $p$-brane expands into the $S^{p+2}$ part of the geometry, 
it has associated a 
maximum angular momentum. As is well-known, these configurations were 
in fact proposed in \cite{GST} as a way to satisfy the stringy exclusion 
principle implied by the AdS/CFT correspondence.

The giant graviton analysis of \cite{GST} was in terms of the Abelian
gauge theory relevant to the description of the expanded brane.
Since massless particles, in particular gravitons, are the source terms 
for gravitational waves, it seems natural to assume that the dielectric 
effect for gravitational waves provides a microscopic picture for giant 
gravitons. Such a microscopical description has so far only been 
given in some particular case. In \cite{DTV} a solution corresponding to 
$N$ M-theory gravitons expanding into an M2-brane of topology 
$\R^{1,1}\times S^2_{NC}$ was
studied in terms of D0-branes expanding into a D2-brane, and then
uplifted to M-theory. This microscopical effect was referred to as a magnetic
moment effect, because the expanded brane carries a magnetic, instead of
an electric, dipole moment.

A second type of giant gravitons, in which the expanded brane has an electric
dipole moment, was considered in \cite{GMT,HHI}. In $AdS_m\times S^n$
spacetimes, these gravitons expand the $AdS$ part of the geometry, and 
therefore do not provide a realisation of the stringy exclusion principle. 
They are referred to in the literature as dual giant gravitons.
Other giant graviton solutions in various backgrounds have been studied
elsewhere in the literature \cite{CR, CR2, BMN}.

In this paper, we aim at describing dielectric configurations from the
point of view of expanding gravitational waves. We will argue that it is 
possible to describe multiple gravitational waves using Matrix theory. As we 
will discuss, Matrix string theory describes, in the
static gauge, coinciding Type IIA gravitational waves. Therefore, taking
the linearised action of Type IIA Matrix string theory in a weakly curved
background, constructed in \cite{schiappa} and \cite{BJL}, we can
identify the extra non-perturbative couplings of non-Abelian gravitons to 
background fields, and use them to study dielectric configurations. 
This same analysis can be done in Type IIB using T-duality. 
In this paper we have constructed the linearised
world volume couplings of multiple gravitons to closed string fields in the
two Type II theories. We have to stress, however, that it would be
necessary to go beyond the linear order calculation presented in this paper
in order to study the giant graviton configurations of \cite{GST}-\cite{HHI},
given that our couplings represent a linear perturbation to the Minkowski 
spacetime, and are therefore not suitable to describe $AdS_m \x S^n$ 
backgrounds. 

We start in section 2 by reviewing in some detail the T-duality relation
between fundamental strings and gravitational waves in the Abelian case,
since this will be of use later. We show here that the
effective action of a gravitational wave can be defined in terms of a 
gauged sigma model
in which the direction of propagation of the wave enters as a special
Killing direction. This action is related through a Legendre transformation
to the usual action for the ten-dimensional massless particle.
In section 3, we present the derivation of the effective 
action for Type IIA multiple gravitational waves from the Matrix string theory 
action. As a check we show that at weak coupling this action reproduces the 
linearised, Abelian effective action derived in section~2.
Some of the couplings in this action were also derived by one of us 
\cite{yolanda}, using various string dualities.
In section 4, a further T-duality takes us to the Type IIB
theory, where we construct the Matrix theory action that describes multiple 
Type IIB gravitational waves. This action is S-duality invariant, as
predicted by the analysis of the Type IIB supersymmetry algebra, and reduces 
also, at weak coupling, to a certain linearised Abelian effective action 
for a massless particle, derived in section 2. 
In section 5, we consider various dielectric solutions that arise from the 
non-Abelian couplings derived in this paper. We also comment on the duality 
relations between these solutions, as well as on
their microscopical picture interpretation for giant 
gravitons. 

\sect{T-duality between fundamental strings and gravitational waves}

We start in this section discussing in some detail
the construction of the effective
action that describes a single gravitational wave. 
It is well known that gravitational
waves and fundamental strings are related by a T-duality transformation
along the direction of propagation of the wave. This works simply at the
level of the supergravity solutions \cite{BEK}. The supergravity solution 
corresponding to a gravitational wave propagating in the $x$ direction,
\be
ds^2= - (2-H)dt^2 + Hdx^2 - 2(1-H)dtdx  + dy_1^2 + ... +dy_{8}^2,
\label{W}
\ee 
is mapped under T-duality along this direction into the supergravity
solution of a fundamental string,
\bea
&&ds^2\  =\ H^{-1} (-dt^2 + dx^2) + dy_1^2 + ... + dy_8^2, \nn [.3cm]
&& e^{-2\phi}\ =\ H, \hspace{2cm} 
B_{0x} \ =\ H^{-1}-1 ,     
\label{F1}
\eea
where $x$ is now the spatial world volume direction of the string.

At the level of the effective actions, the procedure is a little more subtle. 
It is well known that a T-duality transformation applied on the non-linear 
sigma model describing a string propagating in a non-trivial background
yields a new non-linear sigma model describing a string moving in 
the dual background defined by Buscher's rules. 
Instead of performing the most general analysis, however, it is 
possible to start with a very specific string state, a state with zero 
momentum and non-zero winding number, and apply a T-duality transformation. 
The string state thus obtained is a state with only momentum, which can be 
described effectively by a gravitational wave type of action.    

Using this duality connection between string states with winding number and 
string states with momentum, one can simply construct the effective action 
associated to the wave from the Nambu-Goto action for a fundamental string, 
keeping in mind that the action thus derived is only an effective action for a 
specific string state. 
As we show next, one is led in this manner to a gauged sigma model, in which 
the direction of propagation of the wave appears as a special isometric 
direction, in such a way that the gravitational field is transversal to the 
direction of propagation.

The Nambu-Goto action of the string is given in our notation by:
\be
S_{F1}\ =\ -T_1 \int d^2 \sigma\ \sqrt{ | \det 
                  (\part_\alpha X^\mu \part_\beta X^\nu g_\mn) |}
        \ +\ \half T_1 \int d^2\sigma\ \varepsilon^{\alpha\beta} 
                         \part_\alpha X^\mu \part_\beta X^\nu  B_\mn .
\label{Nambu-Goto}
\ee
If we assume that the ninth direction is compact and that the string is 
wound $m$ times around this direction, we can make a split in the target 
space coordinates,
\be
X^\mu (\tau, \sigma) = (X^a (\tau) , X^9(\sigma))=( X^a (\tau) , 
                                 \ m {\mbox{\large $\sigma$}} ),
\ee
and T-dualise along $X^9$. Using Buscher's rules in the kinetic term
of the action we arrive at:
\begin{eqnarray}
&&{\rm det}(\partial_\alpha X^\mu \partial_\beta X^\nu g_{\mu\nu})=
\partial_\tau X^a \partial_\tau X^b (g_{99}g_{ab}-g_{a9}g_{b9}) \nonumber\\
&&\longrightarrow \partial_\tau X^\mu \partial_\tau X^\nu \frac{1}{g_{99}}
(g_{\mu\nu}-\frac{g_{\mu 9}g_{\nu 9}}{g_{99}})\equiv \partial X^\mu
\partial X^\nu {\cal G}_{\mu\nu} .
\end{eqnarray}
Here $\part$ denotes derivation with respect to $\tau$ and ${\cal G}_\mn$ is a 
reduced metric, which we can write as
\be
{\cal G}_\mn = g_\mn  - k^{-2} k_\mu k_\nu,
\ee
introducing a Killing vector pointing along the $X^9$,
or $x$, direction: $k^\mu=\delta^\mu_x$, so that we have
\be
k_\mu = g_{\mu x},
\hspace{2cm}
k^2 = k_x = g_{xx}.
\ee
Similarly, the Wess-Zumino term gives:
\begin{equation}
B_{ax}\rightarrow -\frac{g_{ax}}{g_{xx}}=-\frac{k_a}{k^2}.
\end{equation}
We then arrive at:
\bea
S_W = \ -m  T_0 \int d \tau\  \Bigl\{ k^{-1} \sqrt{ |  
                  \part X^\mu \part X^\nu \ {\cal G}_\mn | }
                 \ \  + \ k^{-2} k_a \part X^a  \Bigr\} . 
\label{Waction}
\eea 
It is straightforward to see that the components of ${\cal G}_\mn$ in 
the $x$ direction vanish,
\be
k^\mu {\cal G}_\mn = 0,
\ee
which states that the gravitational field is transversal to the direction
of propagation of the wave, since we are going to identify soon the
$x$ direction with the direction of propagation.
The reduced metric projects out the world volume scalar $X^9$, thus
removing the world volume degree of freedom 
corresponding to the direction of propagation of the wave.
The action (\ref{Waction}) can in fact be written as a
gauged sigma model, invariant under local transformations along the
$x$ direction:
\bea
S_W = \ -m  T_0 \int d \tau\  \Bigl\{ k^{-1} 
                \sqrt{ | \cD X^\mu \cD X^\nu \ g_\mn | }
                \ \  +  A -\partial X^9 \Bigr\}.  
\label{Wactiongauged}
\eea  
${\cal D}X^\mu$ denote gauge covariant derivatives
\be
\cD  X^\mu \ \equiv \ \part  X^\mu - A  k^\mu 
         \ \equiv \ \part  X^\mu - k^{-2}  k_\rho \part X^\rho k^\mu, 
\label{covder}
\ee
with respect to the scaling symmetry
\be
\delta X^\mu = \Lambda (\tau) k^\mu, 
\hsp{2cm}
\delta A = \part \Lambda (\tau).
\label{scasym}
\ee
Note that the covariant derivatives reduce to ordinary derivatives for 
$\mu \neq x$, while 
for $\mu = x$ we have $\cD X^9 = - k^{-2} k_a \part X^a$.
The action (\ref{Wactiongauged}) is manifestly 
invariant under the transformations (\ref{scasym}).
Gauged sigma models of the type of (\ref{Waction}) and (\ref{Wactiongauged})
have been used before in order to describe the effective 
actions of Kaluza-Klein monopoles \cite{BJO} and M-branes 
in massive eleven-dimensional supergravity \cite{BLO,BvdS}.

The string winding number, $m$, is now the momentum number of the wave 
in the isometry direction, given that
\be
m =\frac{1}{T_0} \frac{\delta S_W}{\delta (\part X^9)},
\ee
a fact that identifies the isometry direction with the direction of
propagation of the wave.

The action (\ref{Wactiongauged}) can in fact be rewritten as the action
of the massless ten-dimensional particle.
In order to see this, we should think
of (\ref{Wactiongauged}) as a Legendre transformed action in which
the dependence on $\partial X^9$ has been replaced by a dependence on its
conjugate momentum $P_9$, but in such a way that a total derivative term
$P_9 \partial X^9 $ is kept in order to have a gauge invariant Lagrangian
\cite{BLO}. 
We then write (\ref{Wactiongauged}) as: 
\bea
S_W[P_9] = \ -\int d \tau\  P_9\ \Bigl\{ k^{-1} 
                \sqrt{ | \cD X^\mu \cD X^\nu \ g_\mn | }
                \ \  +  A -\partial X^9 \Bigr\}.  
\label{W2actiongauged}
\eea  
Integration over $X^9$ fixes $P_9$ to a constant. It is convenient now to 
introduce an auxiliary metric $\gamma$ on the world line and define an action 
$S_W[P_9,\gamma]$ such that (\ref{W2actiongauged}) is recovered upon integration
on $\gamma$:
\bea
S_W[P_9,\gamma] &=& \ -\frac{m  T_0}{2} \int d \tau\  \sqrt{|\gamma|}\
\Bigl\{ \gamma^{-1}  \cD X^\mu \cD X^\nu \ g_\mn + \ k^{-2} 
(\frac{P_9}{mT_0})^2\Bigr\} \nonumber \\ 
&& \hsp{2cm}
+ \int d\tau\ P_9\ (\partial X^9-A).  
\label{Wactiongauged2}
\eea  
{}From this action we easily restore the dependence on $\partial X^9$
through a Legendre transformation:
\be
\label{Legendre}
S[\gamma]=\int d\tau \Bigl\{-P_9 \partial X^9 + L[P_9,\gamma]\Bigr\}, \qquad 
\partial X^9=\frac{\delta L[P_9,\gamma]}{\delta P_9} ,
\ee
in such a way that, after integrating $P_9$, we obtain the action for 
the massless ten-dimensional particle:
\be
\label{masspar}
S[\gamma]=-\frac{mT_0}{2}\int d\tau\ \sqrt{|\gamma|}\ \gamma^{-1}
\partial X^\mu \partial X^\nu g_{\mu\nu} .
\ee

We have thus checked that the action (\ref{Wactiongauged}), T-dual to
the Nambu-Goto action (\ref{Nambu-Goto}), describes a massless particle
in the ten-dimensional spacetime. When the direction of propagation and
the momentum carried by the particle are specified, the more suitable
description is in terms of the action (\ref{Wactiongauged}), in which
the direction of propagation is singled out as a special isometric
direction, and the momentum arises as the charge with respect to the
background field $k_\mu/k^2$.

We now compute for future reference the expansion of the effective action 
(\ref{Wactiongauged}) up to linear order in the background fields. For small
fluctuations in the background
metric $g_\mn \approx \eta_\mn + h_\mn$, we find
\bea
S_W^{\rm lin} &=& -m T_0 \int d\tau \Bigl\{  
1 - \half \dX^2 
- \half h_{00} - \tfrac{1}{4} h_{00} \dX^2 
- h_{0i} \dX^i - \half h_{ij} \dX^i \dX^j   \nonumber\\ [.3cm]
&& 
- \half h_{xx} + \tfrac{1}{4} h_{xx} \dX^2 + 
h_{0x} + h_{ix} \dX^i +\dots \Bigr\}, 
\label{WlinA}
\eea
where the indices $i,j$ run from 1,...,8 $\dot{X}^2\equiv\eta_{ij}
\dot{X}^i \dot{X}^j$ and the dot indicates derivation with respect to the 
time-like coordinate $\tau$.

\begin{figure}
\begin{center}
\leavevmode
\epsfxsize=8cm
\epsffile{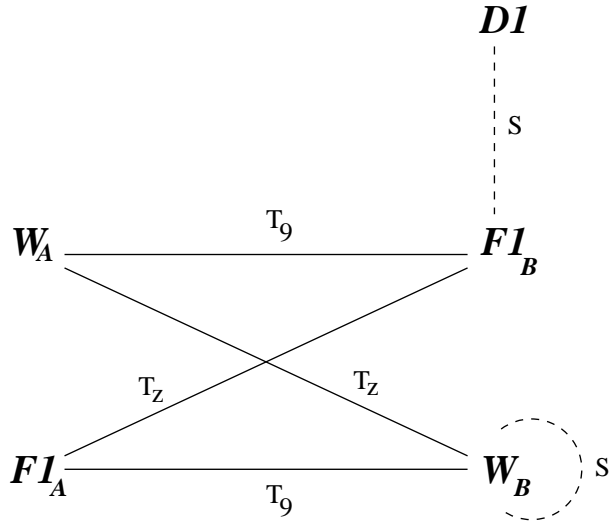}
\caption{\footnotesize The T-duality relations between fundamental strings 
and gravitational waves in the Type II theories. $T_9$ denotes T-duality 
along $X^9$, the world volume direction of the string or propagation 
direction of the wave, while $T_z$ denotes T-duality along a direction 
transverse to the string or the wave. Also shown is the S-duality 
relation between the D-string and the fundamental string of Type IIB and 
the S-duality invariance of the gravitational wave in Type IIB.}  
\label{figuur1}
\end{center}  
\end{figure}

Let us now apply a T-duality transformation in a direction transverse to the 
direction of propagation of the wave that we have just studied.
{}From (\ref{W}) we easily see that the 
supergravity solution remains invariant\footnote{An overview of the 
T-duality relations between gravitational 
waves and fundamental strings can be found in 
Figure \ref{figuur1}.}.
Therefore, we expect that this T-duality transformation, when applied
to the action (\ref{Wactiongauged}), will yield again the action for a 
gravitational wave. In fact, the T-duality
transformation gives rise to a complicated action containing two gauged 
isometries, but which can again be related,
through a Legendre transformation,
to the (dimensional reduction of the) action for the massless ten-dimensional 
particle.
We present this calculation in detail. Indeed, we will see that
the non-Abelian extension
of this action, with two isometries, arises as the Matrix theory
description of multiple Type IIB waves that we will present in section 4. 

Applying T-duality on the action (\ref{Wactiongauged}) along a transverse
direction $Z$ we find:
\bea
S_{W} &=& \ -m T_0 \int d\tau\ \Bigl\{ |l| \ |k^2 l^2 - (k.l)^2 +
(\incl_k\incl_l B)^2|^{-1/2}. \nonumber\\ [.3cm]
&& . \ \sqrt{ |{\cal D}X^\mu {\cal D}X^\nu g_{\mu\nu} +
\frac{k^2-(k.l)^2/l^2}{k^2 l^2-(k.l)^2 + (\incl_k\incl_l B)^2}
\ {\tilde {\cal F}}^2|} \nonumber\\ [.3cm]
&& + \ A^{(1)}-\partial X^9 + 
\frac{\incl_k\incl_l B}{k^2 l^2 - (k.l)^2 + (\incl_k\incl_l B)^2}
\ {\tilde {\cal F}}\Bigr\}.
\label{WBactiongauged}
\eea
This action contains two isometric directions, the direction of propagation
of the original wave, $X^9$, and the transverse direction $Z$ in which we 
performed the T-duality. We have denoted the corresponding Killing vectors 
as $k^\mu=\delta^\mu_x$ as before and $l^\mu=\delta^\mu_z$. The gauge 
covariant derivatives ${\cal D}X^\mu$ are now defined in such a way that 
they gauge the two local isometric transformations generated by $k^\mu$ 
and $l^\mu$
\be
\delta X^\mu = \Lambda^{(1)}(\tau) \ k^\mu + \Lambda^{(2)}(\tau) \ l^\mu,
\ee
that is,
\bea
\label{covderIIB}
{\cal D}X^\mu &\equiv& \partial X^\mu - A^{(1)} k^\mu - A^{(2)} l^\mu
\\ [.3cm]
&\equiv& \partial X^\mu-\frac{l^2(k_\rho \partial X^\rho)-(k\cdot l)
(l_\rho \partial X^\rho)}{k^2l^2-(k\cdot l)^2}\,k^\mu
\ - \frac{k^2(l_\rho \partial X^\rho)-(k \cdot l)(k_\rho \partial X^\rho)}
{k^2l^2-(k \cdot l)^2}\,l^\mu. \nonumber
\eea
${\cal D}X^\mu {\cal D}X^\nu g_{\mu\nu}$ can equivalently be written in
terms of a reduced metric, $\partial X^\mu\partial X^\nu {\cal G}_{\mu\nu}$,
such that its components in the $x$ and $z$ directions vanish:
\be
k^\mu {\cal G}_{\mu\nu}=l^\mu {\cal G}_{\mu\nu}=0.
\ee
This states that the gravitational field is transversal to the two isometric
directions. 
Note that the constant 
$m=\frac{1}{T_0}\frac{\delta S_W}{\delta(\partial X^9)}$
is again the momentum of the wave in the $X^9$ direction, a fact that
identifies the $X^9$ isometric direction with the direction of propagation of
the wave. The other isometric direction, $Z$, is however not physical,
but just an artifact of the T-duality transformation. We are therefore 
describing waves which are smeared in the $Z$-direction. We will see below
that we can in fact rewrite (\ref{WBactiongauged}) as a dimensional
reduction along the $Z$ direction. One can then restore the $Z$ dependence
by simply undoing the reduction. 

On the other hand, ${\tilde {\cal F}}$ is 
an invariant field strength, defined in terms of the T-dual of the transverse 
direction $Z$, that here plays the role of an extra world volume scalar 
\cite{EJL}, and the contraction of the NS-NS 2-form with the Killing
vector $l^\mu$, $(\incl_l B)_\mu=l^\nu B_{\nu\mu}=B_{z\mu}$:
\be
\label{priminv}
{\tilde {\cal F}}\equiv\partial Z + (\incl_l B)_\mu{\cal D}X^\mu.
\ee
Finally, in our notation $\incl_k\incl_l B\equiv B_{zx}$.

The action (\ref{WBactiongauged}) is a complicated expression, which is
however equivalent
to the action for the massless ten-dimensional particle. 
We can see this explicitly by performing a Legendre transformation from
$P_9$ to $\partial X^9$, for which we first rewrite (\ref{WBactiongauged})
as:
\bea
S_{W} [P_9] &=& \ -\int d\tau\ P_9 \Bigl\{ |l| \ |k^2 l^2 - (k.l)^2 +
(\incl_k\incl_l B)^2|^{-1/2}. \nonumber\\ [.3cm]
&&. \ \sqrt{ |{\cal D}X^\mu {\cal D}X^\nu g_{\mu\nu} +
\frac{k^2-(k.l)^2/l^2}{k^2 l^2-(k.l)^2 + (\incl_k\incl_l B)^2}
\ {\tilde {\cal F}}^2|} \nonumber\\ [.3cm]
&& + \ A^{(1)}-\partial X^9 + 
\frac{\incl_k\incl_l B}{k^2 l^2 - (k.l)^2 + (\incl_k\incl_l B)^2}
\ {\tilde {\cal F}}\Bigr\},
\label{WBLeg}
\eea
where integration over $X^9$ fixes $P_9$ to a constant. Now we introduce
an auxiliary metric $\gamma$ on the world line and define the action
$S_W[P_9,\gamma]$ as
\bea
S_{W}[P_9,\gamma] &=& \ -\frac{mT_0}{2} \int d\tau\ \sqrt{|\gamma|}\
 \Bigl\{ \gamma^{-1} \Bigl({\cal D}X^\mu {\cal D}X^\nu g_{\mu\nu} +
\frac{k^2-(k.l)^2/l^2}{k^2 l^2-(k.l)^2 + (\incl_k\incl_l B)^2}
\ {\tilde {\cal F}}^2\Bigr) \nonumber\\ [.3cm]
&& \hsp{3.2cm} + \frac{l^2}{k^2 l^2-(k.l)^2 + (\incl_k\incl_l B)^2}\
(\frac{P_9}{mT_0})^2\Bigr\} \nonumber\\ 
&&+\int d\tau\ P_9\ \Bigl(\partial X^9 - A^{(1)}- 
\frac{\incl_k\incl_l B}{k^2 l^2 - (k.l)^2 + (\incl_k\incl_l B)^2}
\ {\tilde {\cal F}}\Bigr),
\eea
such that (\ref{WBLeg}) is recovered upon integration on $\gamma$. 
The Legendre transformation defined by (\ref{Legendre}) gives rise to:
\be
S[\gamma]=-\frac{mT_0}{2}\int d\tau\ \sqrt{|\gamma|}\ \gamma^{-1}
\Bigl\{{\cal D} X^\mu {\cal D} X^\nu g_{\mu\nu} + l^{-2} (\partial Z +
(i_l B)_\mu \partial X^\mu)^2\Bigr\}, 
\ee
which is the dimensional reduction of the action (\ref{masspar}),
for the ten-dimensional massless particle, along the $Z$ direction, according to 
the T-duality transformed reduction rules \cite{BHO}
\be
g_{zz}=l^{-2},\qquad g_{za}=l^{-2} B_{za},\qquad 
g_{ab}^{(10)}=g_{ab}^{(9)}-l^{-2}B_{za}B_{zb}
\ee
where $\mu=(a,z)$ and $a=0,1,\dots,8$.

We have thus shown that the action (\ref{WBactiongauged}), derived using
T-duality, describes indeed a massless particle, though delocalised in the $Z$
direction. We will see that
multiple Type IIB waves can be described by a non-Abelian extension of
(\ref{WBactiongauged}), but that this action cannot however be related, 
at least
in a simple way, to some non-Abelian extension of (\ref{masspar}),
where the dependence on the isometric directions has been restored.
We will discuss this point further in section 4.

As we did before, we now compute, for future reference, the expansion of 
the action (\ref{WBactiongauged})
up to linear order in the background fields. We find
\bea
\label{WlinB}
&&S_{W_B}^{\rm lin}  =  -m T_0 \int d\tau \Bigl\{  
1 - \half \dX^2 - \half \dZ^2 
- \half h_{00} - \tfrac{1}{4} h_{00} \dX^2
- \tfrac{1}{4} h_{00} \dZ^2 
- h_{0i} \dX^i - \half h_{ij} \dX^i \dX^j \nonumber \\ [.3cm]
&& 
- \half h_{xx} + \tfrac{1}{4} h_{xx} \dX^2 +
\tfrac{1}{4} h_{xx} \dZ^2 
+ h_{0x} + h_{ix} \dX^i + \half h_{zz} \dZ^2
- B_{z0} \dZ + B_{zx}\dZ - B_{zi} \dX^i \dZ+\dots \Bigr\}. 
\nonumber\\
\eea
Here $i,j=1,\dots,7$ and $\dot{X}^2\equiv\eta_{ij}\dot{X}^i\dot{X}^j$.

\sect{Matrix string theory and multiple gravitational waves}

Our starting point for the study of the effective actions associated
to multiple gravitational waves is the action describing the linear couplings 
to background fields of Type IIA Matrix string theory in a weakly curved 
background. This action has been obtained in \cite{schiappa} and \cite{BJL} 
from Matrix theory in a weakly coupled background 
\cite{TVR,TVR2}, by applying a duality chain involving T-duality and 
the 9-11 flip.

In the notation of \cite{BJL}, the Matrix string theory action for weakly 
coupled backgrounds is given by
\bea
S_{{\rm MST}} &=& \frac{1}{2\pi\beta^2} \int d\tau d\sigma 
~\STr \Bigl\{
\tfrac{1}{2}
   \Bigl[ h_{ab} - \eta_{ab} (\phi - \tfrac{1}{2} h_{99} )\Bigr] I_h^{ab} 
\ + 2 B_{a9} I_s^{a9}
\ + \tfrac{1}{2} \Bigl[\phi + \tfrac{1}{2} h_{99} \Bigr] I_h^{99} \nn [.3cm]
&& - \tfrac{1}{2} \Bigl[ \phi 
\ - \tfrac{3}{2} h_{99}\Bigr] I_\phi  
\ + C^{(1)}_a I_h^{a9}
\ - C^{(3)}_{ab9} I_s^{ab} 
\ - C^{(1)}_9 I_0^9 
\ - h_{a9} I_0^a 
\ + 3 B_{ab} I_2^{ab9} 
 \nn [.3cm]
&&
+\ C^{(3)}_{abc} I_2^{abc}  
\ + \tfrac{1}{12} C^{(5)}_{a_1 \ldots a_4 9} I_4^{a_1 \ldots a_4 9}
\ + \tfrac{1}{60} \tilde{B}_{a_1 \ldots a_5 9} I_4^{a_1 \ldots a_5}
\ + \tfrac{1}{48} N^{(7)}_{a_1 \ldots a_69} I_6^{a_1 \ldots a_6 9}
\nn [.3cm]
&& - \tfrac{1}{336} N^{(8)}_{a_1 \ldots a_7 9} I_6^{a_1 \ldots a_7}
\Bigr\},
\label{MSTA}
\eea
where $\beta = R/\sqaprime$, $R$ is the radius of the eleventh dimension 
in Matrix theory, and the ten spacetime indices are split into
$a=(0,...,8)$ and 9, the latter being the direction in which the Matrix 
strings are oriented. 

The currents $I_h$, $I_s$, $I_\phi$ and $I_p$  give the D0-brane couplings to, 
respectively, the metric $h_\mn$, the Kalb-Ramond form $B_\mn$, the dilaton 
$\phi$ and the 
R-R fields $C^{(p+1)}$ \cite{TVR, TVR2}, and can be related to the currents 
appearing in the linear Matrix theory action\footnote{More precise expressions 
for the Matrix theory couplings, to all orders in both derivatives of the 
background fields and the fermionic coordinates, were constructed in
\cite{DNP,Plef} by dimensional reduction of the eleven-dimensional supermembrane
vertex operators.}. However, in (\ref{MSTA}) they couple to different background
fields because of the duality chain involved in the construction of the action 
\cite{schiappa, BJL}. These currents are functions of the $U(N)$ matrix valued 
embedding scalars $X^i$ ($i=1,\dots, 8$), their $U(N)$-covariant derivatives 
$D_\alpha X^i= \part_\alpha X^i + i [A_\alpha, X^i]$, their commutators
$[X^i, X^j]$ and the Born-Infeld 
vector $A_\alpha$ \footnote{We will choose $A_0=0$ throughout the paper, 
at the expense of losing explicit gauge invariance.}.  
We have included in Appendix A the explicit expressions 
for the currents that will be of relevance to this paper. We refer to 
\cite{BJL} for the expressions of all other currents in our notation. 

In the action (\ref{MSTA}), ${\tilde B}$ denotes the NS-NS 6-form potential,
$N^{(7)}$ is the field that couples minimally to the Type IIA Kaluza-Klein 
monopole, and $N^{(8)}$ couples to the so-called $(6,1^2;3)$-brane, an 
``exotic'' 
brane (not predicted by the Type IIA spacetime supersymmetry algebra), first 
mentioned in \cite{BO,Hull,OPR}\footnote{We have included an Appendix which collects 
some properties of these exotic branes of interest to our work. See Appendix B 
for the notation and the duality relations with the branes predicted by the 
spacetime supersymmetry algebras.}. These fields are related to the R-R 6- and 
8-form potentials through the 9-11 flip: 
\be
C^{(6)}_{a_1\dots a_6}\rightarrow N^{(7)}_{a_1\dots a_6 9},
\hsp{2cm}
C^{(8)}_{a_1\dots a_7 9}\rightarrow -N^{(8)}_{a_1\dots a_7 9}.
\ee

In the limit $g_s \rightarrow 0$, the Born-Infeld vector 
and all R-R fields decouple from the action (\ref{MSTA}) and one is forced 
into a space of commuting matrices. In this way light-cone gauge 
string theory in a weakly coupled background is recovered \cite{BJL}. This 
gives weight to the argument that the Matrix string action can indeed be 
derived from Matrix theory in a weakly coupled background 
via a duality chain.

However, a different picture emerges if one goes to the static gauge, 
where one identifies $X^0$ with the time-like world sheet coordinate $\tau$
and $X^9$ with the spatial world sheet coordinate $\sigma$. We also take all 
other embedding scalars independent of $\sigma$, 
\be
X^0 = \tau, \hsp{1.5cm} 
X^9 = \sigma, \hsp{1.5cm} 
\part_\sigma X^a = 0.
\label{staticgauge}
\ee
We will refer to these conditions as static gauge conditions, though the last
one represents in fact an additional truncation\footnote{We thank the referee
for this remark.}.

The action (\ref{MSTA}) is such that
the 9th direction appears as a special isometric
direction on which 
neither the background fields nor the currents depend. This is translated into 
a reduced, $SO(8)$ transverse rotationally invariant action. We can 
however rewrite (\ref{MSTA}), in the static gauge,
in terms of ten-dimensional pull-backs into a one-dimensional 
world volume, by using the techniques of gauged sigma models. If $k^\mu$ is the 
Killing vector pointing along the $X^9$, or $x$, direction,  
$k^\mu=\delta^\mu_x$, we can achieve invariance under the local isometric 
transformations generated by $k^\mu$,
\be
\delta X^\mu=\Lambda(\tau)\,k^\mu,
\ee
by introducing gauge covariant derivatives, as in (\ref{covder})
\be
\cD  X^\mu =  D  X^\mu - k^{-2}  k_\rho D X^\rho k^\mu, 
\label{covder2}
\ee
only now in terms of $U(N)$-covariant derivatives $D X^\mu$.
In the linear approximation, the gauge covariant derivatives (\ref{covder2}) 
reduce to 
${\cal D}X^\mu=D X^\mu$, for $\mu\neq x$ and ${\cal D}X^9=-h_{ax} DX^a$.
Using gauge covariant pull-backs, constructed with these  gauge covariant
derivatives, it is possible to eliminate the pull-back of the isometric
coordinate, and to reproduce the isometric couplings in the action
(\ref{MSTA}) in a manifestly covariant way.
At the linearised level this is easy to see, given that 
${\cal D}X^9$ is proportional to the linear correction to the flat metric 
and therefore does not contribute to the pull-back to linear order. For example,
for a 1-form background field $\Sigma_\mu$ we have
\be
P[\Sigma] = \Sigma_\mu \cD  X^\mu = \Sigma_0 + \beta \Sigma_i \dX^i , 
\ee
with $i=1,\dots,8$. 

Using gauge covariant pull-backs, we are thus able to write the action 
(\ref{MSTA}) in static gauge in a manifestly covariant way, as 
a gauged sigma model, with the gauge covariant derivatives $\cD X^\mu$ gauging 
away the ninth direction, which is interpreted as an isometric 
embedding scalar. 

Filling in the expressions for the R-R currents as given in Appendix A, we
obtain:
\begin{eqnarray}
\label{P1}
C^{(1)}_9 I^9_0 &=& g_s\beta P[\incl_k C^{(1)}]\wedge F, \\ [.3cm]
h_{a9}I^a_0 &=& P [\incl_k h],\\ [.3cm]
B_{ab} I^{ab9}_2 &=& \tfrac{i\beta^2}{3}P[(\incl_X\incl_X)B]\wedge F
-\tfrac{i\beta}{3}P[\incl_{[A,X]}B], \\ [.3cm]
C^{(3)}_{abc} I^{abc}_2 &=& \tfrac{i\beta}{g_s}
P[(\incl_X\incl_X)C^{(3)}], \\ [.3cm]
C^{(5)}_{a_1\dots a_4 9}I^{a_1\dots a_4 9}_4 &=& -\tfrac{6\beta^3}{g_s}
P[(\incl_X\incl_X)^2\incl_k C^{(5)}]\wedge F +\tfrac{12\beta^2}{g_s}
P[(\incl_X\incl_X)\incl_{[A,X]}\incl_k C^{(5)}], \\ [.3cm]
{\tilde B}_{a_1\dots a_5 9}I^{a_1\dots a_5}_4 &=& \tfrac{30\beta^2}{g_s^2}
P[(\incl_X\incl_X)^2\incl_k {\tilde B}], \\ [.3cm]
N^{(7)}_{a_1\dots a_6 9}I^{a_1\dots a_6 9}_6 &=& -\tfrac{8i\beta^4}{g_s^2}
P[(\incl_X\incl_X)^3\incl_k N^{(7)}]\wedge F+\tfrac{24i\beta^3}{g_s^2}
P[(\incl_X\incl_X)^2\incl_{[A,X]}\incl_k N^{(7)}], \\ [.3cm]
\label{P8}
N^{(8)}_{a_1\dots a_7 9}I^{a_1\dots a_7}_6 &=& \tfrac{56i\beta^3}{g_s^3}
P[(\incl_X\incl_X)^3\incl_k N^{(8)}] ,
\end{eqnarray}
where $F = \part A$ ($A\equiv A_\sigma$) and
\be
(\incl_X \incl_X \Sigma)_{\mu_1 ... \mu_p} \equiv X^j X^i  \Sigma_{ij\mu_1 ...
\mu_p},
\hsp{2cm}
(\incl_{[A,X]} \Sigma)_{\mu_1 ... \mu_p} \equiv [A,X^i] \Sigma_{i\mu_1 ...
\mu_p}.
\ee
The inclusion involving the $A$ field is inherited from the 
$D_\sigma X^i$-terms in the pull-backs, of which in the static gauge only the 
commutator part survives. Note also that $k^\mu$ appears explicitly in some 
terms, projecting the background fields on the isometric direction:
\be
(\incl_k \Sigma)_{\mu_1...\mu_p} \equiv k^\rho \Sigma_{\rho\mu_1...\mu_p}
                              =  \Sigma_{x\mu_1...\mu_p}.
\ee
In this manner, when taking the pull-backs to the world volume, the contribution 
of $X^9$ vanishes automatically.

The $\sigma$-component of the Born-Infeld vector $A$ has now obtained the 
interpretation of a world volume scalar. This world volume scalar is associated 
to D0-branes ``ending on the waves''. Indeed, acting with the 9-11 flip on the 
Born-Infeld field strength of the D-string, we obtain the following 
invariant field strength associated to the world volume scalar $A$: 
\begin{equation}
\label{invfi0}
{\cal F} = \partial A + \frac{1}{g_s}P[C^{(1)}]=
F+\frac{1}{g_s}P[C^{(1)}] , 
\end{equation}
since under the 9-11 flip $B_{\mu 9} \rightarrow - C^{(1)}_\mu$
(see \cite{BJL})\footnote{Note that,
although we denote as well the transformed world volume field by $A$, it has
a different gauge transformation rule than the original $A_\sigma=A$,
since the gauge transformation parameters also change under the 9-11 flip.}.
{}From this invariant field strength only $F$ contributes to linear order.

Substituting (\ref{P1})-(\ref{P8}) in the action (\ref{MSTA}) we obtain the
following  expression for the Chern-Simons action:
\bea
\label{MSTCS}
S^{{\rm CS}}_{{W_A}} &=& \frac{1}{\beta^2} \int d\tau 
~\STr \Bigl\{
- g_s \beta\  P [ \incl_k C^{(1)}] \wedge F
\ -  P [ \incl_k h] 
\ + \tfrac{i \beta}{g_s}  P [(\incl_X \incl_X)C^{(3)}]  \nn [.3cm]
&&
+\ i\beta^2 P[(\incl_X \incl_X) B] \wedge F
\ - \ i \beta P[\incl_{[A,X]} B]
\ +\ \tfrac{\beta^2}{2 g_s^2} P [(\incl_X \incl_X)^2 \incl_k \tB ] 
\nn  [.3cm]
&& 
-\ \tfrac{\beta^3}{2 g_s} P [(\incl_X \incl_X)^2 \incl_k C^{(5)} ] \wedge F 
\ +\ \tfrac{\beta^2}{g_s} P[(\incl_X \incl_X) \incl_{[A,X]} \incl_k C^{(5)} ]
\label{Wa} \\  [.3cm]
&& 
-\ \tfrac{i \beta^4}{6 g_s^2} P [ (\incl_X \incl_X)^3 \incl_k N^{(7)}] \wedge F
+\ \tfrac{i \beta^3}{2 g_s^2} P[(\incl_X \incl_X)^2 
                                 \incl_{[A,X]} \incl_k N^{(7)} ]
\nn  [.3cm]
&& 
- \ \tfrac{i \beta^3}{6 g_s^3} P [ (\incl_X \incl_X)^3 i_k N^{(8)} ] \Bigr\}.
\nonumber
\eea

We claim that this action gives the 
Chern-Simons couplings in the world volume effective action of a system
of coincident waves carrying momentum along the $X^9$ 
direction. This action is valid for weakly coupled background fields, since
this is the approximation used in the Matrix theory calculation.
It might seem strange at first sight that Matrix string 
theory can be used to describe gravitational waves. However, recall
that Matrix string theory describes strings 
with non-zero light cone momentum, with the 
winding degrees of freedom entering only non-perturbatively in the 
description, and that string states with non-zero momentum  
are effectively described in terms of massless particles. It is then not 
surprising that the Matrix string action provides, in the static 
gauge, an effective description for multiple gravitational waves.

We have a number of arguments to support this claim. First of all, the 
monopole term $P[\incl_k h]$ in (\ref{Wa}) is just the linearised expression 
for $A-\partial X^9$ in (\ref{Wactiongauged}), indicating that we are dealing 
with waves with a momentum in the $X^9$ direction. As we have reviewed in 
section~2, waves carrying momentum in a compactified direction are T-dual 
to fundamental strings wound around this direction. Indeed, in \cite{BJL} 
the non-Abelian action describing multiple Type IIB fundamental strings
with non-zero winding number was derived
from our action (\ref{MSTA}) using T-duality along the
$X^9$ direction, which we claim is the direction of propagation of the
waves. 
Let us also stress that we find agreement between   
(\ref{Wa}) and the terms in the action for multiple gravitational 
waves that were constructed in \cite{yolanda}. These observations support
our claim that Type IIA Matrix strings reduce to multiple 
waves in the static gauge.

There are also more quantitative arguments supporting this identification. 
While in the limit 
$g_s \rightarrow 0$, (\ref{MSTA}) in the light-cone frame reduces to the action 
of light-cone gauge string theory in weakly coupled background fields 
\cite{BJL}, in the static gauge the limit $g_s \rightarrow 0$ reduces 
(\ref{MSTA}) to the linearised effective action of a set of commuting 
gravitational waves. The only non-vanishing currents, up to order $\beta^2$,  
are 
\be
\begin{array}{ll}
I_h^{00} = \unity + \half \beta^2  \dX^2, \hspace{2cm} &
I_\phi= \unity - \half \beta^2  \dX^2 ,  \\ [.3cm]
I_h^{0i} = \beta \dX^i,   &
I_0^0 = \unity,  \\ [.3cm]
I_h^{ij} = \beta^2 \dX^i \dX^j,  &
I_0^i = \beta \dX^i,  
\end{array}
\ee
and (\ref{MSTA}) reduces to
\bea
S_{W_A}^{\rm linear} &=& \frac{1}{\beta^2} \int d\tau \Bigl\{
\Bigl[ \half h_{00} - h_{0x} + \half h_{xx} \Bigr] \unity 
+ \beta \Bigr[(h_{0i} - h_{ix})\dX^i\Bigr] 
\nn[.3cm]
&& \hsp{2cm}
+ \tfrac{1}{2} \beta^2 \Bigl[ h_{ij}\dot{X}^i\dot{X}^j +
\tfrac{1}{2} (h_{00} - h_{xx}) \dX^2  \Bigr] +{\cal O}(\beta^3)\Bigr\}.
\eea 
This action coincides, up to the relative rescalings between the Abelian
and Matrix theory calculations, with the linear order contribution to 
the expansion of the effective action (\ref{Wactiongauged}),
for Abelian 
gravitational waves, given by (\ref{WlinA}).

Finally, let us point out that it is also possible to take the Abelian 
limit without taking 
$g_s\rightarrow 0$, in which case we get linearised non-perturbative 
corrections to the action (\ref{Wactiongauged}). Substituting the 
expressions for the NS-NS currents (see Appendix A) in (\ref{MSTA}), 
together with (\ref{P1}), we find:
\begin{equation}
\label{correc}
S_{W_A}^{\rm corr}=\frac{g_s}{\beta}\int d\tau 
\Bigl\{ P[\incl_k C^{(1)}]\wedge F
+\beta C^{(1)}_i \dot{A} \dot{X}^i\Bigr\} .
\end{equation}
One can easily see that the Legendre
transformation of section 2 restores the $X^9$ dependence also in these
correcting terms, giving rise in (\ref{masspar}) to an extra linear coupling
\be
S[\gamma]=-\frac{mT_0}{2}\int d\tau\ \sqrt{|\gamma|}\ \gamma^{-1}
\Bigl\{\partial X^\mu \partial X^\nu g_{\mu\nu} - \frac{g_s}{2}
C^{(1)}_\mu \partial X^\mu\ F\Bigr\}.
\ee
Identifying the full couplings whose linear expansion gives rise to this
contribution is not easy without the help of some additional
requirement. We will discuss in the next section that it is possible 
in principle to 
identify some of these couplings by
imposing S-duality invariance for Type IIB waves, and then T-dualising.

We have thus discussed that the Matrix theory calculation describes Type IIA
waves in terms of a non-Abelian extension of the action (\ref{Wactiongauged}),
for
which the direction of propagation of the waves appears as a special
isometric direction. One could try now to restore the dependence on $X^9$,
by performing a Legendre transformation along the lines of section 2.
This procedure, however, does not work, as it can easily be argued.
First, the direction of propagation of the waves, as it arises in the
calculation, is not matrix-valued, since it is identified with the
spatial world sheet direction of D-strings (see \cite{BJL}).
So, even though we can in principle restore some explicit dependence
on $\partial X^9$, the new terms are going to be Abelian. Moreover,
it is clear that
the Legendre transformation cannot give rise to non-Abelian commutators
of the transverse scalars involving $X^9$. This results in an action with,
still, $SO(8)$ transverse rotational invariance.

In fact, the existence of a compact isometric direction gives the spacetime
on which the waves propagate a non-trivial topology. This compact
direction allows $p$-branes to wind around it or live in its transverse
space, giving rise to the specific dielectric couplings in (\ref{MSTCS}),
for which the $k^\mu$ dependence cannot be eliminated.
These extra topologically non-trivial couplings are in a sense analogous to 
the supergravity solutions of \cite{ET}, corresponding to $p$-branes 
wound around different compact dimensions.


\sect{Multiple Type IIB gravitational waves}

In this section we present the action associated to multiple
Type IIB gravitational waves.
We calculate it by performing a T-duality transformation, along a transverse 
direction, on the Matrix string theory action with linear 
couplings to closed string fields, given by (\ref{MSTA}). 
We argue that this action describes Type IIB waves in 
the static gauge since, as we have seen, Matrix string theory describes
Type IIA waves in this gauge, and we are T-dualising along a transverse
direction (see Figure 1). The action thus obtained contains the linear
couplings of multiple Type IIB gravitons to closed string fields.

\subsection{The action for multiple Type IIB gravitational waves}

Since we are working with actions which are only linear in the background
fields,
it is sufficient to use the T-duality transformation rules to linear order. 
For a T-duality transformation in a generic $y$ direction, these
are
given by:
\be
\label{Tdualidades}
\begin{array}{lll}
h_{yy} \rightarrow - h_{yy},   \hspace{1cm}&
B_{ab} \rightarrow B_{ab}, \hspace{2.3cm}&
N^{(7)}_{a_1\dots a_6 y}  \rightarrow \tB_{a_1\dots a_6},
\\ [.3cm]
h_{a y} \rightarrow -B_{a y}, &
\tB_{a_1\dots a_5 y}  \rightarrow \tB_{a_1\dots a_5 y}, &
(i_k N^{(7)})_{a_1\dots a_5 y} \rightarrow 
(i_k N^{(7)})_{a_1\dots a_5 y},  
\\ [.3cm]
h_{ab} \rightarrow h_{ab}, &
\tB_{a_1\dots a_6}  \rightarrow {\cal N}^{(7)}_{a_1\dots a_6 y}, &
(i_k N^{(7)})_{a_1\dots a_6} \rightarrow 
(i_k {\cal N}^{(8)})_{a_1\dots a_6 y}, 
\\ [.3cm]
\phi  \rightarrow \phi -\half h_{yy}, &
C^{(p)}_{a_1\dots a_{p-1}y}  \rightarrow  C^{(p-1)}_{a_1\dots a_{p-1}}, &
(i_k N^{(8)})_{a_1\dots a_6 y} \rightarrow
(i_k N^{(8)})_{a_1\dots a_6 y}, 
\\ [.3cm]
B_{a y} \rightarrow - h_{a y}, &
C^{(p)}_{a_1\dots a_p} \rightarrow C^{(p+1)}_{a_1\dots a_p y}, &
(i_k N^{(8)})_{a_1\dots a_7} \rightarrow
(i_k N^{(9)})_{a_1\dots a_7 y}. 
\end{array}
\ee
We see that for higher order background fields T-duality gives rise to
new gravitational fields, $N^{(p)}$ and ${\cal N}^{(q)}$, which have in fact
already been encountered in the literature \cite{EJL, EL}. We have summarised 
the notation and the duality properties of these fields in Appendix B. 
The fields $N^{(7)}$ and ${\cal N}^{(7)}$ couple minimally
to Type IIB Kaluza-Klein monopoles with different Taub-NUT directions,
as we will clarify later on in this section. $N^{(8)}$, ${\cal N}^{(8)}$
and $N^{(9)}$ couple to different ``exotic'' branes, not predicted by the
Type IIB spacetime supersymmetry algebra\footnote{A more detailed 
explanation of the role played by these branes is contained in Appendix B.}.

Applying T-duality to the action (\ref{MSTA}) in a transverse
direction $Z$ we obtain the following linear action:
\bea
\label{WBlineal}
S^{{\rm linear}}_{W_B} &=& \frac{1}{2 \pi \beta^2} \int d\tau d\sigma
 ~\STr \Bigl\{
\tfrac{1}{2} \Bigl[h_{ab} - \eta_{ab} (\phi
                -\tfrac{1}{2} h_{zz}-\tfrac{1}{2} h_{99} ) \Bigr] I_h^{ab}
\ -  B_{az} I_h^{az}  \nn [.3cm]
&& \hsp{-1.5cm}
\ -  \tfrac{1}{2} \Bigl[\phi+ \tfrac{1}{2} h_{zz}- \tfrac{1}{2} h_{99}  \Bigr]
I_h^{zz}  
\ +  \ C^{(2)}_{az}I_h^{a9}
\ +  \tfrac{1}{2} \Bigl[\phi - \tfrac{1}{2}h_{zz} + \tfrac{1}{2}h_{99}\Bigr]
I_h^{99} 
\ +  C^{(0)} I_h^{z9} 
  \nn [.3cm]
&& \hsp{-1.5cm}
\ -  \tfrac{1}{2} \Bigl[ \phi - \tfrac{1}{2} h_{zz}  - \tfrac{3}{2} h_{99}\Bigr]
I_\phi 
\ + \ C^{(4)}_{abz9} I_s^{ab} 
+ 2 C^{(2)}_{a9} I_s^{az} 
\ +  2 B_{a9} I_s^{a9} 
\ + 2 h_{z9} I_s^{z9} 
\label{WBlin}
 \\ [.3cm]
&& \hsp{-1.5cm}
\ - h_{a9} I_0^a
\ - B_{z9} I_0^z 
\ + C^{(2)}_{z9} I_0^9
\ + \ C^{(4)}_{abcz} I_2^{abc} 
\ + 3 C_{ab} I_2^{abz}   
\ + 3 B_{ab} I_2^{ab9} 
\ - 6 h_{az} I_2^{az9}
\nn [.3cm]
&& \hsp{-1.5cm}
\ + \tfrac{1}{60} {\cal N}^{(7)}_{a_1 ... a_5 9z} I_4^{a_1 ... a_5}
\ + \tfrac{1}{12} \tB_{a_1 ... a_4 z9} I_4 ^{a_1 ... a_4z}
\ - \tfrac{1}{12} C^{(6)}_{a_1 ... a_4z9} I_4^{a_1 ... a_4 9}
\ - \tfrac{1}{3} C^{(4)}_{abc9} I_4^{abcz9}
\nn [.3cm]
&& \hsp{-1.5cm}
\ - \tfrac{1}{336} N^{(9)}_{a_1 ... a_79z} I_6^{a_1 ... a_7}
\ + \tfrac{1}{48} N^{(8)}_{a_1 ... a_69z} I_6^{a_1 ... a_6z} 
\ + \tfrac{1}{48}{\cal N}^{(8)}_{a_1 ... a_69z} I_6^{a_1 ... a_69}
\ + \tfrac{1}{8} N^{(7)}_{a_1 ... a_5z9} I_6^{a_1 ... a_5z9} \Bigr\}.
\nonumber
\eea
Here now, the indices $a$ run from 0 to 7. As a consistency check we
have 
derived the same action by performing a T-duality transformation in the action 
describing multiple Type IIA fundamental strings, in the static gauge,
constructed in \cite{BJL}, 
this time along the spatial world volume direction of the 
strings (see Figure 1). We recover (\ref{WBlineal}) up to a sign difference
for the R-R fields. Note, however, that this sign difference is not physical.
T-duality transformations in two distinct directions only commute 
up to a sign in the right moving R-sector \cite{Polchinski} 
$(T_1 \cdot T_2 = (-)^{F_R} \ T_2 \cdot T_1)$, which  changes the sign of
all the R-R fields. However, $(-)^{F_R}$ is a symmetry of the theory, and this 
makes this sign difference unphysical.

In analogy with the linear action (\ref{MSTA}), we can argue that the action 
(\ref{WBlin}) describes multiple light-cone fundamental strings in a weakly 
coupled Type IIB background. Our main interest here, however, lays in the 
wave interpretation of the action (\ref{WBlin}), so that we take, instead, 
the static gauge (\ref{staticgauge}). 

In this gauge we obtain a one-dimensional action with two isometries: 
one corresponding to translations along the direction of propagation 
$X^9$ of the waves and the other to translations along the transversal 
direction $Z$, in which the T-duality has been performed. The Matrix theory 
calculation seems then to be describing Type IIB waves in terms of a 
non-Abelian 
extension of the action (\ref{WBactiongauged}), as we will further check. 
In this action the direction of propagation of the waves appears as an 
isometric direction, but there is as well an extra isometry in the $Z$
direction. We are therefore dealing with
multiple gravitational waves which are delocalised in the $Z$ direction. 
As we discussed in section 2, the isometry in $Z$ is however not physical,
but just an artifact of the T-duality transformation. Indeed,
for the Abelian case we showed how it 
is possible to restore the $Z$-dependence in the 
action. However, the corresponding Legendre
transformation in the non-Abelian case
yields still an action in which $X^9$ and $Z$ appear
as special directions (see the discussion at the end of the previous
section). We are thus constrained to work, in the non-Abelian case, 
with an action with two
isometries, and assume the presence of the extra, unphysical isometry. 

As we did in the previous section, we can rewrite the Matrix theory
action (\ref{WBlineal}) in a manifestly covariant way as a gauged sigma
model.
Introducing Killing vectors associated to the two isometries
$k^\mu=\delta^\mu_x$ and $l^\mu=\delta^\mu_z$ and gauge covariant 
derivatives, we can achieve invariance under local isometric 
transformations generated by $k^\mu$ and $l^\mu$:
\be
\delta X^\mu=\Lambda^{(1)}(\tau)\ k^\mu + \Lambda^{(2)} (\tau)\ l^\mu,
\ee
with the gauge covariant
derivatives defined as in (\ref{covderIIB}):
\begin{equation}
{\cal D}X^\mu=DX^\mu-\frac{l^2(k_\rho D X^\rho)-(k\cdot l)
(l_\rho D X^\rho)}{k^2l^2-(k\cdot l)^2}\,k^\mu
\ - \frac{k^2(l_\rho D X^\rho)-(k \cdot l)(k_\rho D X^\rho)}
{k^2l^2-(k \cdot l)^2}\,l^\mu,
\end{equation} 
now in terms of $U(N)$-covariant derivatives $DX^\mu$.
In the linear approximation, ${\cal D}X^\mu=D X^\mu$ for $\mu\neq x,z$, 
whereas ${\cal D}X^9=-h_{ax}D X^a$ and ${\cal D}Z=-h_{az}D X^a$. 
Therefore, at the linearised level, ${\cal D}X^9$ and ${\cal D}Z$
do not contribute to the pull-backs.

Filling in the expressions for the R-R currents as given in Appendix A,
we can then write the Chern-Simons
part of (\ref{WBlin}), in the static gauge, in a manifestly covariant
way, as:

\bea
S^{{\rm CS}}_{{W_B}} &=& \frac{1}{\beta^2} \int d\tau 
~\STr \Bigl\{ 
-  P [\incl_k h] 
\ - ig_s\beta [A,\omega] P[\incl_l h] 
\ - ig_s\beta^2 P[\incl_{[\omega,X]} \incl_l h]\wedge F
\nn [.3cm] 
&& \hsp{-1.5cm}%
\ + ig_s\beta^2 P[\incl_{[A,X]}\incl_l h ]\wedge {\tilde F} 
\ + g_s\beta P[\incl_l\incl_k B]\wedge {\tilde F} 
\ - i\beta P[\incl_{[A,X]} B]
\ + i\beta^2 P[(\incl_X \incl_X)B]\wedge F 
\nn [.3cm]
&& \hsp{-1.5cm}%
\ - g_s \beta P[\incl_l\incl_k C^{(2)}]\wedge F 
\ - i\beta P[\incl_{[\omega,X]}C^{(2)}] 
\ + i\beta^2 P[(\incl_X\incl_X)C^{(2)}]\wedge {\tilde F} 
\ - \tfrac{i\beta}{g_s}P[(\incl_X\incl_X)\incl_l C^{(4)}] 
\nn [.3cm]
&& \hsp{-1.5cm} %
\ + \beta^2 P[\incl_{[A,X]}\incl_{[\omega,X]}\incl_k C^{(4)}] 
\ - \beta^2[A,\omega] P[(\incl_X\incl_X)\incl_k C^{(4)}]  
\ - \beta^3 P[\incl_{[A,X]}(\incl_X\incl_X)\incl_k C^{(4)}]
                 \wedge {\tilde F} 
\nn [.3cm]
&& \hsp{-1.5cm} %
\ - \beta^3 P[\incl_{[\omega,X]} (\incl_X\incl_X)i_k C^{(4)}]
                                \wedge F
\ + \tfrac{\beta^2}{g_s}P[\incl_{[A,X]}(\incl_X\incl_X)\incl_l\incl_k C^{(6)}]
\ - \tfrac{\beta^3}{2g_s}P[(\incl_X\incl_X)^2\incl_l\incl_k C^{(6)}]\wedge F 
\nn [.3cm]
&& \hsp{-1.5cm}%
\ - \tfrac{\beta^2}{g_s}P[\incl_{[\omega,X]}(\incl_X\incl_X)\incl_l\incl_k\tB] 
\ + \tfrac{\beta^3}{2g_s}P[(\incl_X\incl_X)^2\incl_l\incl_k \tB]\wedge \tF 
\ - \tfrac{\beta^2}{2g_s^2}P[(\incl_X\incl_X)^2\incl_l\incl_k {\cal N}^{(7)}] 
\nn [.3cm]
&& \hsp{-1.5cm}%
\ + \tfrac{i\beta^3}{g_s}P[\incl_{[A,X]} (\incl_X\incl_X)
                       \incl_{[\omega,X]}\incl_l\incl_k N^{(7)}] 
\ - \tfrac{i\beta^3}{2g_s}[A,\omega] P[(\incl_X\incl_X)^2 
                       \incl_l\incl_k N^{(7)}] 
\nn [.3cm]
&& \hsp{-1.5cm}%
\ + \tfrac{i\beta^4}{2g_s} P[(\incl_X\incl_X)^2\incl_{[\omega,X]}
                      \incl_l\incl_k N^{(7)}] \wedge F 
\ - \tfrac{i\beta^4}{2g_s}P[(\incl_X\incl_X)^2 \incl_{[A,X]}
                      \incl_l\incl_k N^{(7)}]\wedge {\tilde F} 
\label{WBCS}
\\ [.3cm] 
&& \hsp{-1.5cm}%
\ + \tfrac{i\beta^3}{2g_s^2}P[\incl_{[A,X]}(\incl_X\incl_X)^2 
                      \incl_l\incl_k {\cal N}^{(8)}] 
\ - \tfrac{i\beta^4}{6g_s^2} P[(\incl_X\incl_X)^3
                      \incl_l\incl_k {\cal N}^{(8)}]\wedge F 
\nn[.3cm]%
&& \hsp{-1.5cm}
\ + \tfrac{i\beta^3}{2g_s^2}P[\incl_{[\omega,X]}(\incl_X\incl_X)^2
                      \incl_l\incl_k N^{(8)}] 
\ - \tfrac{i\beta^4}{6g_s^2} P[(\incl_X\incl_X)^3
                      \incl_l\incl_k N^{(8)}]\wedge {\tilde F}
\ + \tfrac{i\beta^3}{6g_s^3}P[(\incl_X\incl_X)^3\incl_l\incl_k N^{(9)}]\Bigl\}
.
\nonumber
\eea
Like the action for Type IIA waves, this action gives the Chern-Simons
couplings of a system of coincident Type IIB waves for weakly coupled
background fields, due to the linear approximation of the Matrix theory
calculation.

Here $F=\partial A$ and ${\tilde F}=\partial\omega$, where we have redefined
the coordinate $Z$, which now plays the role of 
a world-volume scalar \cite{EJL}, as 
\begin{equation}
Z\equiv g_s \omega .
\end{equation}
This rescaling will be useful when we study the behaviour of the action
under S-duality.
The commutator inclusion terms are defined as
\be
(\incl_{[A,X]} \Sigma)_{\mu_1 ... \mu_p} \equiv [A,X^i] 
\Sigma_{i\mu_1 ... \mu_p},
\hsp{2cm}
(\incl_{[\omega,X]} \Sigma)_{\mu_1 ... \mu_p} 
       \equiv [\omega,X^i] \Sigma_{i\mu_1 ... \mu_p},
\ee
and $\incl_l \incl_k \Sigma$ denotes double contractions:
\begin{equation}
(\incl_l \incl_k \Sigma)_{\mu_1\dots\mu_p}
\equiv l^\lambda k^\rho \Sigma_{\rho\lambda\mu_1\dots\mu_p}
=\Sigma_{xz\mu_1\dots\mu_p}.
\end{equation}   
Background fields contracted in this manner are such that both the $X^9$ 
and $Z$ contributions vanish when taking the pull-backs to the world volume . 

Note that, as in the Type IIA case, the monopole term  $P[\incl_k h]$ in 
the Chern-Simons action above is the origin of the momentum carried
by the waves 
in the propagation direction $X^9$.

The action for multiple Type IIB gravitational waves that we have thus 
derived seems 
particularly involved, compared to the dielectric action of other $p$-branes. 
This is because we are dealing with an action that has two compact isometry 
directions, which give the spacetime we consider a non-trivial topology. The 
two compact directions allow $p$-branes to wind around them, giving rise to 
the different inclusion terms that characterise the action (\ref{WBCS}).

Note that there are two different gravitational 7-forms that couple in the
action (\ref{WBCS}), $N^{(7)}$ and ${\cal N}^{(7)}$, which emerge through 
the T-duality rules (\ref{Tdualidades}). These fields can couple in the
Chern-Simons action (\ref{WBCS}) only because there exist two isometric
directions. $N^{(7)}$ is electric-magnetic dual to the Killing vector
$k_\mu$, considered as a 1-form, and ${\cal N}^{(7)}$ to $l_\mu$.
Therefore, they are associated to KK-monopoles 
with Taub-NUT directions pointing along the two different Killing vectors. 
Similarly, $N^{(8)}$ and ${\cal N}^{(8)}$ 
are Poincar\'e duals to the scalars  $k^2$ and $l^2$
(see Appendix B).

We will now give some arguments that support the interpretation of the 
action (\ref{WBCS}) as the non-Abelian Chern-Simons action for multiple 
Type IIB gravitational waves. First we will show the S-duality invariance, 
not only of (\ref{WBCS}), but of the full action (\ref{WBlin}),
in agreement with the predictions of the Type IIB supersymmetry algebra.
Secondly we will show that, in analogy with the 
Type IIA case, we obtain in the weak coupling limit, the linearised 
Abelian action for a massless particle.

\subsection{S-duality invariance}
\label{invariance}

In this section we check that the action that we have proposed for multiple 
gravitational waves in linearised Type IIB backgrounds is manifestly S-duality 
invariant. This is a non-trivial check for our action, given that the 
S-selfduality of the (single) Type IIB gravitational wave is predicted by the 
analysis of the spacetime supersymmetry algebra of the theory \cite{Hull2}. 
The fact that the invariance also holds for the non-Abelian case is remarkable
and an argument in favour of our interpretation.

To linear order, the S-duality rules for the metric, the dilaton and the R-R 
scalar are given by
\be
h_\mn \rightarrow h_\mn - \eta_\mn \phi, \hsp{2cm}
 \phi \rightarrow - \phi, \hsp{2cm}
C^{(0)}   \rightarrow -C^{(0)} ,
\label{bulkS1}
\ee 
while the 2-form, 6-form and 8-form fields combine in S-duality doublets 
\be
\begin{array}{lll}
B \rightarrow -C^{(2)},    \hsp{1cm} 
& {\tilde B} \rightarrow -C^{(6)},     \hsp{1cm} 
& {\cal N}^{(8)} \rightarrow -N^{(8)}, \\ [.3cm] 
C^{(2)} \rightarrow B, 
& C^{(6)} \rightarrow {\tilde B},
& N^{(8)} \rightarrow {\cal N}^{(8)}.
\label{bulkS2}
\end{array}
\ee
These transformation rules for $N^{(8)}$ and ${\cal N}^{(8)}$ can easily
be derived taking into account the origin of these fields
from eleven dimensions, where S-duality
is realised as a modular transformation in the 2-torus. Indeed, 
we easily see from Figure \ref{exotics} in Appendix B, that both 
fields have their M-theory origin in the field that couples minimally to the 
Kaluza-Klein monopole, and that one gets one or the other by simply 
interchanging the eleventh and T-duality directions, which is 
well known to generate S-duality in the Type IIB theory \cite{BHO}.

$N^{(7)}$, ${\cal N}^{(7)}$ and $N^{(9)}$ are on the other hand
singlets. This is clear for $N^{(7)}$, ${\cal N}^{(7)}$,
since they couple to Type IIB Kaluza-Klein monopoles with different
Taub-NUT directions, and these are known to be S-duality invariant
\cite{Hull2}. One can check it explicitly by looking at the gauge 
transformation rules of these fields, computed in \cite{EJL}.
That this is also the case for $N^{(9)}$ was shown in \cite{EL}.

Let us now study the S-duality transformation rules of the
world volume fields. The two world volume fields $A$ and $\omega$ form 
a doublet under S-duality, as we can deduce by looking at the invariant 
field strengths they form with the pull-backs of the background fields. 
T-dualising the invariant field strength (\ref{invfi0}), that couples in 
the world volume of the Type IIA waves, we derive the following invariant 
field strength for the world volume scalar $A$:
\be
\label{invfi1}
{\cal F}=\partial A-\frac{1}{g_s} P[\incl_l C^{(2)}].
\ee
The invariant field strength for $\omega$ was in turn derived in 
section 2, for the Abelian case. Expression (\ref{priminv})
becomes $g_s \partial\omega + (\incl_l B)_\mu {\cal D}
X^\mu$. Therefore in our Matrix theory notation we have:
\be
\label{invfi2}
{\tilde {\cal F}}=\partial\omega + \frac{1}{g_s} P[\incl_l B].
\ee
We thus see that $A$ and $\omega$ are associated to D-strings and
F-strings respectively, wound around the $Z$ direction, ending on the
waves. Given that $B$ and $C^{(2)}$ transform as a doublet under S-duality 
(see (\ref{bulkS2})), $\omega$ and $A$ must transform as:
\be
\omega \rightarrow A, \hsp{2cm}
A \rightarrow  -\omega .
\label{wvS}
\ee

With this behaviour of the background and world volume fields under 
S-duality, it is straightforward to check that the Chern-Simons action 
(\ref{WBCS}) is indeed invariant under S-duality.

As to the Born-Infeld part of the action (\ref{WBlin}), it is shown in
Appendix A that, at least up to order $\beta^2$, the matrix currents 
$(I_h^{a9},I_h^{az})$ and $(I_s^{a9},I_s^{az})$ form doublets 
under the transformations (\ref{wvS}), and that $I_h^{9z}$ picks up a 
minus sign, while $I_h^{99}$ and $I_h^{zz}$ get mapped into each other. 
All other currents are left invariant. 

It is not difficult to see that the Born-Infeld action, at least up to order 
$\beta^2$, is then S-duality invariant.

\subsection{The Abelian limit}

As in the previous section, we can support our claim that the current
Matrix theory calculation describes Type IIB gravitational waves by
taking the weak coupling limit and checking that the action (\ref{WBlineal}) 
reduces to the linearised effective action for a set of commuting Type
IIB gravitational waves. As we already discussed, Matrix theory describes
Type IIB waves in terms of a non-Abelian extension of the action
(\ref{WBactiongauged}), so we should compare, in the weak coupling limit,
with the linear expansion of this action, given by expression
(\ref{WlinB}).
 
Again, in the limit $g_s\rightarrow 0$, the Born-Infeld vector and all 
but the NS-NS fields decouple, and we are forced into a space of commuting 
matrices. The expressions for the relevant currents can be found in Appendix A.
In this limit they are given, up to order $\beta^2$, by
\be
\begin{array}{ll}
I_\phi = \unity -\half \beta^2
               \Bigl(\dX^2 + \dZ^2\Bigr), 
\hsp{2.5cm} & I_h^{iz} = \beta^2 \dX^i \dZ, \\ [.3cm]
I_h^{00} = \unity + \half \beta^2 
               \Bigl(\dX^2 + \dZ^2\Bigr),&
I_h^{zz} = \beta^2 \dZ^2 ,\\ [.3cm]
I_h^{0i} = \beta \dX^i , &
I_0^0 = \unity, \\ [.3cm]
I_h^{0z} = \beta \dZ , & I_0^i =  \beta \dX^i ,\\ [.3cm]
I_h^{ij} = \beta^2 \dX^i \dX^j,&
I_0^z =  \beta \dZ,  
\label{gs0currents}
\end{array}
\ee
with all other currents equal to zero.
The action (\ref{WBlin}) then reduces to
\bea
\label{linearIIBgs0}
S_{W_B}^{\rm linear} &=& \frac{1}{\beta^2} \int d\tau \Bigl\{
\Bigl[ \half h_{00} - h_{0x} + \half h_{xx} \Bigr] \unity 
+ \beta \Bigr[(h_{0i} - h_{ix})\dX^i-(B_{0z}+B_{zx})\dot{Z} \Bigr] 
\nn[.3cm]
&& \hsp{-1cm}
+ \tfrac{1}{2} \beta^2 \Bigl[ h_{ij}\dot{X}^i\dot{X}^j +
\tfrac{1}{2} (h_{00} - h_{xx}) (\dX^2 + \dZ^2) - h_{zz}\dot{Z}^2
-2 B_{iz}\dot{X}^i\dot{Z} \Bigr]+{\cal O}(\beta^3) \Bigr\},
\eea
where $\dot{X}^2\equiv \eta_{ij}\dot{X}^i\dot{X}^j$.
Therefore we reproduce, up to rescalings,
the linearised effective action (\ref{WlinB})
for commuting gravitational waves derived in section 2.

If we take the Abelian limit without sending the string coupling constant
to zero, we recover the action (\ref{WBactiongauged})
supplemented with non-perturbative corrections
associated to D-brane degrees of freedom. These corrections are indeed
necessary in order to achieve invariance of (\ref{WBactiongauged}) 
under S-duality, as predicted by the Type IIB supersymmetry algebra. 
{}From the linearised Matrix theory calculation we find
the non-perturbative corrections:
\bea
\label{linearIIBgsn0}
S_{W_B}^{\rm corr}&=&\frac{g_s}{\beta} \int d\tau  
\Bigl\{ - \incl_l\incl_k C^{(2)} \wedge F 
        - P[\incl_l C^{(2)}] \wedge F + \beta C^{(0)} F \dot{Z} 
\\
&& \hsp{2cm}
        + g_s \beta \Bigl[ \phi F^2 
                           + \tfrac{g_s \beta}{4}
                   (h_{00}-h_{xx}-2h_{zz}) F^2  \Bigl] \Bigl\}.
\nonumber
\eea
Note that the two terms in $C^{(2)}$ are precisely the S-duals of the
$B$ terms in (\ref{linearIIBgs0}). Indeed
it is easy to check that the sum of (\ref{linearIIBgs0}) and
(\ref{linearIIBgsn0}) is S-duality invariant.

We can derive this S-duality invariant action
as the linear expansion of the
following non-perturbative modification of (\ref{WBactiongauged}):
\bea
S_{W_B} &=& \ -m T_0 \int d\tau\ \Bigl\{ |l| \ |k^2 l^2 - (k.l)^2 +
(\incl_k\incl_l B)^2|^{-1/2}. \nonumber\\ [.3cm]
&&. \ \sqrt{ |{\cal D}X^\mu {\cal D}X^\nu g_{\mu\nu} +
\frac{k^2-(k.l)^2/l^2}{k^2 l^2-(k.l)^2 + (\incl_k\incl_l B)^2}
\ e^\phi\ {\rm F}^T {\rm M} {\rm F}|} \nonumber\\ [.3cm]
&& + \ A^{(1)}-\partial X^9 + 
\frac{\incl_k\incl_l B \wedge {\tilde F}-\incl_k\incl_l C^{(2)} \wedge F}
{k^2 l^2 - (k.l)^2 + (\incl_k\incl_l B)^2} \Bigr\}.
\label{WBactiongauged2}
\eea
Here ${\rm F}^T {\rm M} {\rm F}$ is the $SL(2,\R)$-invariant combination:
\begin{equation}
\label{SL2Rinv}
{\rm F}^T {\rm M} {\rm F} = \left( \begin{array}{cc}{\tilde {\cal F}} 
& {\cal F} \end{array}
\right)  e^\phi \left( \begin{array}{cc}
e^{-2 \phi} + C^{(0)\, 2} & C^{(0)} \\
C^{(0)} & 1 \end{array} \right)
\left( \begin{array}{c} {\tilde {\cal F}} \\ {\cal F}
\end{array} \right)
\, ,
\end{equation}
with
\be
\label{otrainv}
{\tilde {\cal F}}=\partial Z + P[\incl_l B],\qquad
{\cal F}=\partial A - P[\incl_l C^{(2)}],
\ee
the gauge invariant curvatures of the world volume scalars $Z$ and
$A$. 
Note that ${\tilde {\cal F}}$ is the field strength that
coupled in the perturbative action (\ref{WBactiongauged}) (see 
(\ref{priminv})), and that 
${\cal F}$ is its S-dual, which has to couple as well in the non-perturbative
action in order to achieve invariance under S-duality\footnote{The relative
$1/g_s$ factor between (\ref{otrainv}) and the field strengths given in
(\ref{invfi1}) and (\ref{invfi2}) is due to the rescalings of the 
Matrix theory calculation, that take 
$\partial A \rightarrow g_s \partial A$ (see Appendix A).}.
The $SL(2,\R)$-invariant combination (\ref{SL2Rinv}) appears as well in the
world volume effective action of the
Type IIB Kaluza-Klein monopole \cite{EJL}, a
brane which is also known to behave as a singlet under S-duality.
 
This extension of (\ref{WBactiongauged2}) is, however,
not fully S-duality invariant, since we have not included the S-dual
of the $(\incl_k\incl_l B)^2$ term. Since this term 
does not contribute to the linearised action, we cannot determine
the correct S-duality invariant combination associated to it by comparing 
with the Matrix theory calculation. By T-dualising the full S-duality 
invariant 
action we would be able to derive as well the non-perturbative
couplings in the action of Abelian Type IIA waves, whose linear expansion
we determined using Matrix theory.

\sect{Dielectric wave solutions}

In this section we present some explicit solutions of gravitational 
waves expanding into dielectric branes, rising from the non-Abelian couplings 
derived in the previous sections.

We have seen in sections 3  and 4 that the Matrix theory calculation captures, 
already to linear order, the non-perturbative degrees of freedom associated 
to D-brane states. These new  degrees of freedom translate into correction 
terms, (\ref{correc}) and (\ref{linearIIBgsn0}), to the perturbative, Abelian 
actions (\ref{Wactiongauged}) and (\ref{WBactiongauged}). 
In the non-Abelian case we find as well non-perturbative terms corresponding to 
the contraction of the background fields with the embedding scalars, yielding 
multipole couplings of the same type as those derived in \cite{myers, TVR2} for 
the action of coinciding D-branes. These couplings are responsible for 
the expansion of the waves into dielectric configurations, in the presence 
of external field strengths. 

In \cite{schiappa} it was shown that the term 
\begin{equation}
\label{diele}
\frac{i}{\beta g_s}P[(\incl_X\incl_X)C^{(3)}]
\end{equation}
in the action (\ref{Wa}), gives rise to a potential
\be
V(X) = \STr \Bigl\{- 
       \frac{1}{4 g_s^2} [X, X]^2 + \frac{i}{3 g_s} X^k X^j X^i F^{(4)}_{0ijk} 
            \Bigr\},
\label{Wpotential}
\ee
which has a stable solution of the form
\be
F^{(4)}_{0ijk} = f \epsilon_{ijk}, 
\hsp{2cm}
X^i = - \tfrac{1}{4} g_s f J^i,
\label{W-D2}
\ee
where the matrices $J^i$ form an $N \x N$ representation of $SU(2)$ 
\cite{myers}.

The interpretation of the Matrix string action as describing multiple 
Type IIA waves helped clarifying this Matrix string dielectric solution 
(see section 7.3 in \cite{BJL}). The solution in \cite{schiappa} was 
interpreted in terms of $N$ fundamental strings expanding into a 
fuzzy D2-brane, with world volume $\tau$ times a 2-sphere in 
the $(i,j,k)$-directions. However, the world volume of the D2-brane
does not contain the spatial world sheet direction 
$X^9$ of the string, nor can the dipole coupling $(\incl_X\incl_X) C^{(3)}$ 
in the Chern-Simons action appear as a pull-back to a two-dimensional 
world volume, as required in the expanding strings interpretation. 
Schiappa pointed out that the expanding fundamental strings were actually
describing gravitons, though no explicit check of this assessment was given.

With our interpretation of Matrix string theory as describing multiple
gravitational waves we can see explicitly that this solution does indeed
correspond to multiple gravitons expanded into a D2-brane, which is in
fact transverse to their direction of propagation\footnote{Recall that 
the pull-backs are 
defined in terms of covariant derivatives  (\ref{covder2}).} (see also 
\cite{BJL}). Moreover, within this interpretation (\ref{diele}) can naturally 
be written as a pull-back to the one-dimensional world line of the waves. 

Other terms in the action (\ref{Wa}) describe multipole couplings of the waves 
to NS5-branes and KK-monopoles, although, to our knowledge, no stable 
dielectric solutions associated to these terms are known.

Similarly, in the Type IIB action (\ref{WBCS}) the dielectric coupling 
\begin{equation}
\label{diele2}
\frac{i}{\beta g_s}P[(\incl_X\incl_X) \incl_lC^{(4)}]
\end{equation}
gives rise to a stable non-Abelian solution of the form
\be
F^{(5)}_{0zijk} = f \epsilon_{ijk}, 
\hsp{2cm}
X^i = - \tfrac{1}{4} g_s f J^i,
\label{W-D3}
\ee
which we interpret as $N$ Type IIB gravitational waves expanding into a 
D3-brane transverse to the propagation direction $X^9$ of the waves, 
and wrapped around the $Z$ direction. The rest of dielectric couplings
in the action (\ref{WBCS})
describe multipole couplings to KK-monopoles and exotic branes. 

These gravitational wave configurations represent a
microscopical description
of the dual giant gravitons of \cite{GMT, HHI}, where we have however ignored
the 
back reaction of the geometry.\footnote{Recall that one has to go beyond 
the linear approximation presented in this paper in order to describe properly 
giant gravitons in an 
$AdS_m \x S^n$ space.}

\begin{figure}
\begin{center}
\leavevmode
\epsfxsize=8cm
\epsffile{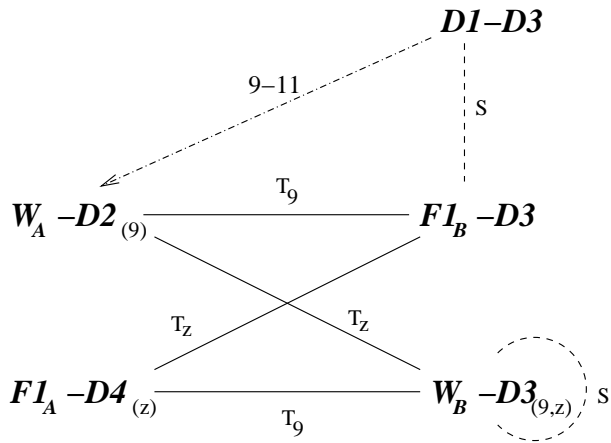}
\caption{\footnotesize Duality relations between  
solutions corresponding to dielectric fundamental strings and 
gravitational waves in the
Type IIA and Type IIB theories. In our notation $W_A - D2$ represents
multiple Type IIA waves expanded into a D2-brane.
As in Figure \ref{figuur1}, $T_9$ denotes a 
T-duality transformation in the propagation direction $X^9$ of the waves, 
or world sheet spatial direction of the strings, while $T_z$ indicates a 
T-duality transformation in a direction orthogonal to the waves and the 
strings. Also shown is the S-duality between the D1-D3 
solution and the F1-D3 solution in Type IIB, the S-duality invariance 
of Type IIB gravitational waves expanding into a D3-brane and the 
9-11 flip that maps the D1-D3 solution into expanding Type IIA 
gravitational waves. The lower indices between brackets indicate the 
isometry directions of the corresponding world volume theory.}  
\label{solutions}
\end{center}  
\end{figure}
If we take into account the non-Abelian character of the actions for 
multiple fundamental strings and multiple 
waves that we have presented here and in our previous paper \cite{BJL}, 
it is remarkable how well the duality relations hold, both for the 
actions as for their solutions. Dielectric solutions describing fundamental 
strings expanding into D3 and D4-branes have been given in \cite{BJL}.  
These solutions are T-dual to the solutions that we have just discussed
for dielectric gravitational 
waves, just as one may expect by naively applying the duality relations
shown in Figure \ref{solutions}. Also, 
S-duality acting on the Type IIB 
solutions gives the expected answer, as can be seen in this Figure: 
the F1-D3 configuration is S-dual to the well known D1-D3 solution, and 
the gravitational wave expanding into a D3-brane is clearly S-duality 
invariant. 
 
There exists as well another type of polarisation effect, referred to
as the magnetic 
moment effect \cite{DTV}, where $p$-branes with a non-zero velocity polarise
under 
the influence of an external magnetic field (as opposed to an electric field,
for the dielectric effect). In this reference it was shown that a set of $N$ 
D0-branes in an external magnetic flux tube, with a non-zero velocity 
along the flux tube, can be described by a Lagrangian
\be
L= \STr \Bigl\{ \half  \dX^2 +  \frac{1}{4 g_s^2} [X, X]^2 
               + \frac{i}{6 g_s} \dX^l X^k X^j X^i F^{(4)}_{ijkl} 
            \Bigr\} ,
\label{magnpotential}
\ee
where $\dot{X}$ indicates 
differentiation with respect to the world line coordinate $\tau$.
This Lagrangian admits a solution of the form
\be
F^{(4)}_{ijkl} = f \epsilon_{ijkl}, 
\hsp{1.5cm}
X^a = \tfrac{1}{4} g_s f v J^a, 
\hsp{1.5cm}
X^4 = v \tau \unity,
\label{W-D2v}
\ee
where the coordinates $i$ are split as $i = (a,4)$, $X^4$ is the direction
on which the D0-branes move, and, 
as in (\ref{W-D2}), the matrices $J^a$ form an $N \x N$ representation 
of $SU(2)$. 
This solution describes D0-branes expanding into a 
D2-brane orthogonal to the direction $X^4$ in which the D0-branes move. 
The radius of the D2-brane turns out to be independent of $N$, but 
proportional to the momentum of the D0-branes, at least for $N \gg 1$.
This solution 
was compared to a spherical D2-brane (with the 2-sphere in the 
$(a,b,c)$-directions)
embedded in a magnetic R-R flux tube and moving in 
the $X^4$ direction, and it was found that both the radius and the energy 
of the D2-brane coincide with those provided by the microscopical picture,
in terms of expanding D0-branes. As also discussed in \cite{DTV}, the 
eleven-dimensional interpretation of (\ref{W-D2v}) should be in terms of 
gravitons, moving both along the eleventh direction and along $X^4$,  
polarising into a spherical M2-brane. Therefore, this solution provides
a microscopical description of a particular eleven-dimensional giant graviton
configuration.

Following the construction of \cite{DTV} 
it is not difficult to see that the dielectric coupling (\ref{diele}) for 
multiple Type IIA gravitational waves gives rise to the same Lagrangian  
(\ref{magnpotential}), now with gauge covariant pull-backs into the world
line.
A solution identical to (\ref{W-D2v}) will therefore 
also describe ten-dimensional gravitational waves moving in the $X^9$ and 
in the $X^4$ directions, expanding into a D2-brane 
transverse to the $X^9$ direction. 
This is just the 
dimensionally reduced version of the eleven-dimensional picture
of \cite{DTV}, as we will further discuss in the conclusions.

Reference \cite{DTV} considered as well a solution corresponding to
eleven-dimensional gravitons propagating only in the $X^9$ direction.
The corresponding Lagrangian is similar to 
(\ref{magnpotential}), but with a magnetic field strength with 
$F^{(4)}_{ijk9}$ components. Reduction over $X^9$ yields a system 
of D0-branes in a magnetic NS-NS field $H_{ijk}$,  
where the coupling arises from the Born-Infeld, rather 
than from the Chern-Simons action (see \cite{myers}). 
It would be interesting to construct 
the analogous version of this solution in terms of Type IIA 
gravitational waves in a $F^{(4)}_{ijk9}$ flux tube. However, due to the 
fact that in our approach the propagation direction $X^9$ of the 
waves is isometric, it is not clear how to obtain from the action (\ref{MSTA})
a coupling of the form
\be
\STr \{ \dX^9 X^k X^j X^i  F^{(4)}_{ijk9} \},
\ee 
a difficulty that should be related to the problem, that we have already
discussed, of restoring the dependence in the 
isometric direction in the non-Abelian case.

\sect{Conclusions}

We have provided some evidence that Matrix string theory describes, in the 
static gauge, coincident Type IIA gravitational waves. We have given some 
weight to this identification by showing that Matrix string theory in static 
gauge and in a weakly coupled background reproduces in the $g_s\rightarrow 0$ 
limit, the linearised action describing Abelian Type IIA gravitational waves 
in a weakly coupled background, and by showing agreement with previous results
derived from string dualities \cite{yolanda}.

The action thus proposed contains the linear couplings of non-Abelian
gravitons to closed string fields. These terms include non-Abelian,
Myers-like couplings that are responsible for the expansion of the
gravitons into higher-dimensional branes. We have studied some particular
configurations, corresponding to multiple gravitational waves expanding into
a spherical D2-brane, in the presence of an external electric flux tube, and 
in a  spherical D2-brane with non-zero transverse velocity, in the presence 
of an external magnetic flux tube. We related these solutions to similar 
configurations in eleven dimensions \cite{DTV}.

Perhaps the most interesting dielectric effect associated to gravitational
waves is the one giving rise to the giant graviton configurations in
$AdS_m\times S^n$ backgrounds. We would need however to go beyond the
linear order approximation of this paper in order to provide a microscopical
description of these configurations, given that a linear perturbation
of a flat background is not an adequate framework in which to describe the
$AdS$ background. 
It should however be possible to study giant graviton configurations
in the Penrose limit, given that the pp-wave background can be 
approximated by a linear perturbation to Minkowski\footnote{An analysis of 
eleven-dimensional gravitational waves 
with dielectric couplings in a pp-wave background, arising as the Penrose 
limit 
of $AdS_4 \x S^7$ and $AdS_7 \x S^4$, has been given in \cite{BMN}.}, or
in more general backgrounds, like those presented in \cite{DTV, CR, CR2}.

We have also provided a Matrix theory description for non-Abelian
gravitons in the Type IIB theory. The key ingredient is the T-duality
connection between Type II waves, that we have applied to the Matrix
string theory action in static gauge in a weakly coupled background.
The action thus obtained reproduces in the $g_s\rightarrow 0$ limit a
particular linearised action describing Abelian Type IIB gravitational
waves. A non-trivial check of this action is its S-duality invariance,
a property that had been predicted by the analysis of the Type IIB
spacetime supersymmetry algebra \cite{Hull2}. We have also given a non-Abelian 
solution of Type IIB gravitons expanding into a dielectric D3-brane. 

Finally, we should stress an important point about the actions for 
gravitational
waves that we have derived. If we start with the Matrix theory action
describing Type IIA waves and perform a T-duality transformation along
the direction of propagation of the waves, we arrive at a Matrix theory
formulation of Type IIB fundamental strings in the static gauge \cite{BJL}, 
to which we have already referred in this article. 
As shown in \cite{BJL}, this action is actually S-dual to the
world volume effective action describing multiple D-strings, a fact that
suggests that this Matrix description of Type IIB F-strings, and
therefore the one here proposed to describe Type IIA waves, 
are perhaps only suitable to describe 
the corresponding branes in the strong coupling regime.

Moreover, the Matrix theory description of Type IIA waves is in terms
of the non-Abelian extension of a particular description for a  
ten-dimensional massless
particle, in terms of an action in which the direction of
propagation of the particle plays a special role (it is isometric).
This action is related to the usual description of the ten-dimensional 
massless
particle by means of a Legendre ($\leftrightarrow$ world volume duality)
transformation. Now, world volume duality transformations
typically map the weak and the strong coupling regimes of a theory.
Therefore, it is possible that the dielectric configurations that 
we have studied in section 5 are 
suitably described by the Myers-like couplings that we have computed
in this paper only in the strong coupling regime. Still, dielectric
configurations of this type are expected to occur as well
in the weak coupling regime, at least those preserving some fraction of
the supersymmetries, though perhaps they are described microscopically
in terms of different variables. It would be quite interesting to
further study whether this is really the case.

One remarkable aspect of
our description of non-Abelian Type IIA waves is that it can be used to
describe non-Abelian gravitons in M-theory. Indeed, since it adequately
describes Type IIA gravitons in the strong coupling regime, we can
describe eleven-dimensional gravitons by just
rewriting ten-dimensional fields in terms of eleven-dimensional 
ones. The coupling (\ref{diele}):
\be
\frac{i}{\beta g_s}P[(\incl_X\incl_X)C^{(3)}],
\ee
that we used to construct the dielectric solution associated to Type
IIA waves expanded into a D2-brane, does actually agree with the dipole
coupling for eleven-dimensional waves predicted in \cite{yolanda}.
This solution is in fact the dimensional reduction of $N$ 
eleven-dimensional gravitational 
waves polarised into a 
spherical M2-brane, with the 2-sphere in the $(i,j,k)$-space, where the
reduction has to be performed 
in a direction orthogonal to both the direction of propagation
and $X^i$, $X^j$ and $X^k$. 
$N$ M-theory gravitons expanded in a spherical M2-brane have been
considered in \cite{BMN}, in the pp-wave background. 
The off-diagonal terms in the pp-wave metric give a mass to the
embedding
scalars in the non-Abelian action, and the solution can be made
supersymmetric by an appropriate choice of the mass. Dielectric
solutions
with non-zero masses were considered first in \cite{TV}.
On the other hand, a reduction of the eleven-dimensional 
solution along the 
propagation direction $X^9$ of the waves yields the well-known solution 
associated to 
$N$ D0-branes expanding into a dielectric D2-brane, already studied by 
Myers in \cite{myers}. In fact,
a nice check of the eleven-dimensional action for the M-theory gravitons
would be to reproduce the linearised action for multiple D0-branes
\cite{TVR}
when reducing along the direction of propagation of the gravitons.

It would also prove interesting to derive the action for M-theory
gravitons directly in a Matrix theory set-up, i.e. using Matrix theory
in a weakly coupled background \cite{KT,TVR1}. A 
Matrix theory description of M-theory gravitons in
a pp-wave background has recently been given in 
\cite{BMN,CLP,DSR}\footnote{See \cite{bonelli} for a formulation of
Matrix string theory in a pp-wave background.}.
This Matrix model is determined by the supersymmetry
algebra of the pp-wave background, and consists on the usual Matrix
theory of BFSS \cite{BFSS} plus a correcting term with
a dielectric coupling which is shown to
be responsible for the expansion of the gravitons into a fuzzy M2-brane.
It would be interesting to try to make contact with these results within
our description. We hope to report progress in 
this direction in a forthcoming publication \cite{JL}.

\vspace{1cm}
\noindent
{\bf Acknowledgements}\\

\vsp{-.3cm}
\noindent
We wish to thank Dominic Brecher, Roberto Emparan, Clifford Johnson,
Nakwoo Kim, 
Tom\'as \Ortin , M.A.R. Osorio and Dave Page for the 
useful discussions. B.J.~thanks the Instituto de F\'{\i}sica Te\'orica of the 
U.A.M./C.S.I.C.~in Madrid and the Departamento de F\'{\i}sica of the Universidad 
de Oviedo for their hospitality during part of this work. 
Y.L. would also like to thank the Institute for Theoretical Physics in Leuven
and the AEI in Potsdam for hospitality.

\vspace{1cm}
\appendix

\renewcommand{\theequation}{\Alph{section}.\arabic{equation}}

\sect{Matrix  currents}

In this Appendix we include, for completeness, the explicit expressions 
for the currents that couple in the gravitational wave actions that we have 
constructed. Here we have taken static gauge (\ref{staticgauge}) as 
compared to \cite{TVR}. This results in the fact that all terms proportional 
to $D_\sigma X$ reduce to terms proportional to $[A, X]$.  Note also that our
currents differ from the ones in \cite{TVR}, due to the fact that we use 
dimensionless world sheet coordinates
\be
\sigma = \frac{g_s}{\sqaprime}\  x, 
\hspace{2cm} 
\tau = \frac{R}{\aprime} \ t,
\label{eqn:rescalings}
\ee
and have performed the rescalings $X^i \longrightarrow \sqrt{\aprime} X^i$ and 
$g_s \longrightarrow g_s/(2\pi)$.  This results in practise in a rescaling of 
the currents, with respect to \cite{TVR}, by a factor of $\beta = R/\sqaprime$ 
for each term of the form $\dX$, $[X,X]$, $[A,X]$ and $\dA$, and a factor of 
$g_s$ and $1/g_s$ for each $\dA$ and $[X,X]$ respectively. 

\subsection{R-R currents}

With $i,j=1,\ldots,8$, and up to order $\beta^2$, we have, for the 
Type IIA waves:
\bea
I_0^0 &=& \unity, \nn 
I_0^i &=& \beta\dot{X}^i, \nn 
I_0^9 &=& g_s \beta \dot{A}, \nn 
I_2^{0ij} &=& - \tfrac{i\beta}{6 g_s}  [X^i, X^j], \nn 
I_2^{09i} &=& -\tfrac{i\beta}{6} [A,X^i], \nn 
I_2^{ijk} &=& -\tfrac{i \beta^2}{6 g_s} 
               \left( \dot{X}^i [X^j,X^k] + \dot{X}^j [X^k,X^i] 
                      + \dot{X}^k [X^i,X^j] \right), \nn 
I_2^{9ij} &=& -\tfrac{i\beta^2}{6} 
                \left( \dot{A} [X^i, X^j] - \dot{X}^i [A,X^j]
                       + \dot{X}^j [A,X^i] \right), \nn 
I_4^{0ijkl} &=& -\tfrac{\beta^2}{2g_s^2} 
                 \left( [X^i,X^j][X^k,X^l] + [X^i,X^k][X^l,X^j] 
                        + [X^i,X^l][X^j,X^k] \right), \nn 
I_4^{09ijk} &=& -\tfrac{ \beta^2}{2 g_s} 
                \left( [A,X^i] [X^j,X^k] + [A,X^j] [X^k,X^i] 
                       + [A,X^k] [X^i,X^j] \right), \nn 
I_4^{ijklm} &=& -\tfrac{15 \beta^3}{2g_s^2} 
                 \dot{X}^{[i} [X^j,X^k][X^l,X^{m]} ], \\
I_4^{9ijkl} &=& -\tfrac{3 \beta^3}{2 g_s}  
                 \left( \dot{A} [X^{[i},X^j][X^k,X^{l]}] 
                        - 4  \dot{X}^{[i} [X^j, X^k] [A,X^{l]}]\right), 
\nn  
I_6^{0ijklmn} &=& \tfrac{i \beta^3}{g_s^3} 
                   [X^{[i},X^j][X^k,X^l][X^m,X^{n]}], \nn 
I_6^{09ijklm} &=& \tfrac{i\beta^3}{g_s^2} 
                   [A,X^{[i}] [X^j, X^k][X^l,X^{m]}], \nn 
I_6^{ijklmnp} &=& \tfrac{7i \beta^4}{g_s^3} 
                   \dot{X}^{[i} [X^j,X^k][X^l,X^m][X^n,X^{p]}], \nn 
I_6^{9ijklmn} &=& \tfrac{i\beta^4}{g_s^2} 
                   \left( \dot{A} [X^{[i},X^j][X^k,X^l][X^m,X^{n]}] 
                         - 6 \dot{X}^{[i} [X^j, X^k][X^l,X^m] [A,X^{n]}] 
                         \right).
\nonumber
\eea
To obtain the currents that couple in the linear action 
(\ref{WBlin}) for Type IIB gravitational waves, one of the transverse 
coordinates, say $X^8$, must be taken to be $Z \equiv g_s \omega$. 

\subsection{NS-NS currents}

The NS-NS currents that couple in the action for Type IIA gravitational
waves can simply be obtained from the Matrix string currents in \cite{BJL}
going to static gauge. 
The corresponding NS-NS currents for Type IIB gravitational
waves have however singled out the $Z$ direction, and we will include
their explicit expressions for the sake of clarity. They
are given, up to
order $\beta^2$, by:
\bea
I_{\phi}& =& \unity - \tfrac{\beta^2}{2} \dot{X}^2
                    - \tfrac{\beta^2}{4g_s^2} [X,X]^2 
                    - \tfrac{g_s^2 \beta^2}{2} ( F^2  + \tF^2 ) 
                    - \tfrac{\beta^2}{2}( [A,X]^2 + [\omega,X]^2 )
                    - \tfrac{g_s^2 \beta^2}{2} [A,\omega]^2,
\nn [.2cm]
I_h^{00} &=& \unity + \tfrac{\beta^2}{2} \dot{X}^2
                     - \tfrac{\beta^2}{4 g_s^2} [X,X]^2
                    + \tfrac{g_s^2 \beta^2}{2} ( F^2+ \tF^2)
                    - \tfrac{\beta^2}{2}( [A,X]^2 +  [\omega,X]^2)
                    - \tfrac{g_s^2 \beta^2}{2} [A,\omega]^2  , 
\nn [.2cm]
I_h^{0i} &=& \beta \dot{X}^i , 
\nn [.2cm]
I_h^{0z} &=& g_s \beta \tF, 
\nn [.2cm]
I_h^{09} &=& g_s\beta F,
\nn [.2cm]
I_h^{ij} &=& \beta^2\dot{X}^i \dot{X}^j
            - \tfrac{\beta^2}{g_s^2} [X^i,X^k][X^k,X^j] 
            + \beta^2 [A,X^i][A, X^j] 
            + \beta^2 [\omega, X^i][\omega,X^j], 
\nn [.2cm]
I_h^{iz} &=& g_s\beta^2\dot{X}^i \tF
            + \tfrac{\beta^2}{g_s}[X^i,X^j]  [\omega, X^j]
            + g_s \beta^2 [A,X^i][A, \omega],
\nn [.2cm]
I_h^{i9} &=& g_s \beta^2  \dot{X}^i F 
           + \tfrac{\beta^2}{g_s} [X^i,X^j] [A, X^j]
           -  g_s \beta^2 [\omega, X^i] [A, \omega], 
\nn [.2cm]
I_h^{zz} &=& g_s^2\beta^2 \tF^2
            + \beta^2 [\omega, X]^2
            + g_s^2\beta^2 [A, \omega]^2,
\nn [.2cm]
I_h^{99} &=& g_s^2 \beta^2 F^2 
           + \beta^2 [A,X]^2 + g_s^2 \beta^2 [A,\omega]^2, 
\\ [.2cm]
I_h^{z9} &=& g_s^2 \beta^2 F \tF 
           + \beta^2 [\omega,X^j] [A, X^j],
\nn [.2cm]
I_s^{0i} &=& - \tfrac{i\beta^2}{2g_s}[X^i,X^j] \dot{X}^j
             +\tfrac{i g_s \beta^2}{2}  [A,X^i] F 
             + \tfrac{ig_s\beta^2}{2}[\omega, X^i] \tF, 
\nn [.2cm]
I_s^{0z} &=& - \tfrac{i\beta^2}{2}[\omega,X^j] \dot{X}^j
             +\tfrac{i g_s^2 \beta^2}{2} F  [A,\omega]  , 
\nn [.2cm]
I_s^{09} &=& -\tfrac{i\beta^2}{2} \dot{X}^i [A,X^i]
             -\tfrac{ig_s^2\beta^2}{2} \tF [A,\omega], 
\nn [.2cm]
I_s^{ij} &=& \tfrac{i\beta}{2g_s} [X^i,X^j] , 
\nn [.2cm]
I_s^{zi} &=& \tfrac{i\beta}{2} [\omega, X^i] , 
\nn [.2cm]
I_s^{9i} &=& \tfrac{i\beta}{2} [A,X^i],
\nn [.2cm]
I_s^{9z} &=& \tfrac{ig_s\beta}{2} [A,\omega],
\nonumber
\eea
where $Z= g_s \omega$, $F=\partial A$, ${\tilde F}=\partial \omega$, and
\begin{eqnarray}
{[}X,X{]}^2 &\equiv& [X^i,X^j][X^i,X^j] , \\
{[}A,X{]}^2 &\equiv& [A,X^i][A,X^i], \\
{[}\omega,X{]}^2 &\equiv & [\omega,X^i][\omega,X^i].   
\end{eqnarray}
It is straightforward to check that $I_\phi$, $I_h^{ab}$, $I_s^{ab}$ and
$I_s^{9z}$ are singlets under the S-duality transformations (\ref{wvS}), 
while $(I_h^{a9},I_h^{az})$ and $(I_s^{a9}, I_s^{az})$ transform as 
doublets. $I_h^{z9}$ changes sign, and $I_h^{99}$ and $I_h^{zz}$ get 
mapped into each other. As shown in subsection \ref{invariance}, these currents
combine in the linear action (\ref{WBlin}) with the background fields in such a 
way that the action is invariant under S-duality transformations.   


\sect{Gravitational background fields and duality}

In this Appendix we summarise the duality properties of the gravitational
fields that arise in the different duality transformations that have
been carried out in the paper, and briefly summarise previous results
in the literature concerning the role they play as sources for
gravitational branes in Type II. We refer the reader to \cite{EL}
for a more extensive study.

Tables \ref{table1} and \ref{table2}
summarise the Type II gravitational branes that are charged w.r.t. these
fields. Some of these branes are exotic branes, in the sense that they 
are not predicted by the standard Type II spacetime supersymmetry algebras 
(see however \cite{ET}). 
Exotic branes have been encountered in the 
literature in different contexts. On the one hand their existence is
required in order to fill up multiplets of BPS states in representations
of the U-duality groups $E_7(\Z)$, $E_8(\Z)$
of M-theory on $T^7$, $T^8$ tori \cite{EGKR,BO,Hull,OPR, OP}.
On the other hand they are predicted by dualities on branes that do occur
as central charges in the spacetime supersymmetry algebras \cite{EL}.

\begin{figure}[!ht]
\begin{center}
\leavevmode
\epsfxsize= 14cm
\epsffile{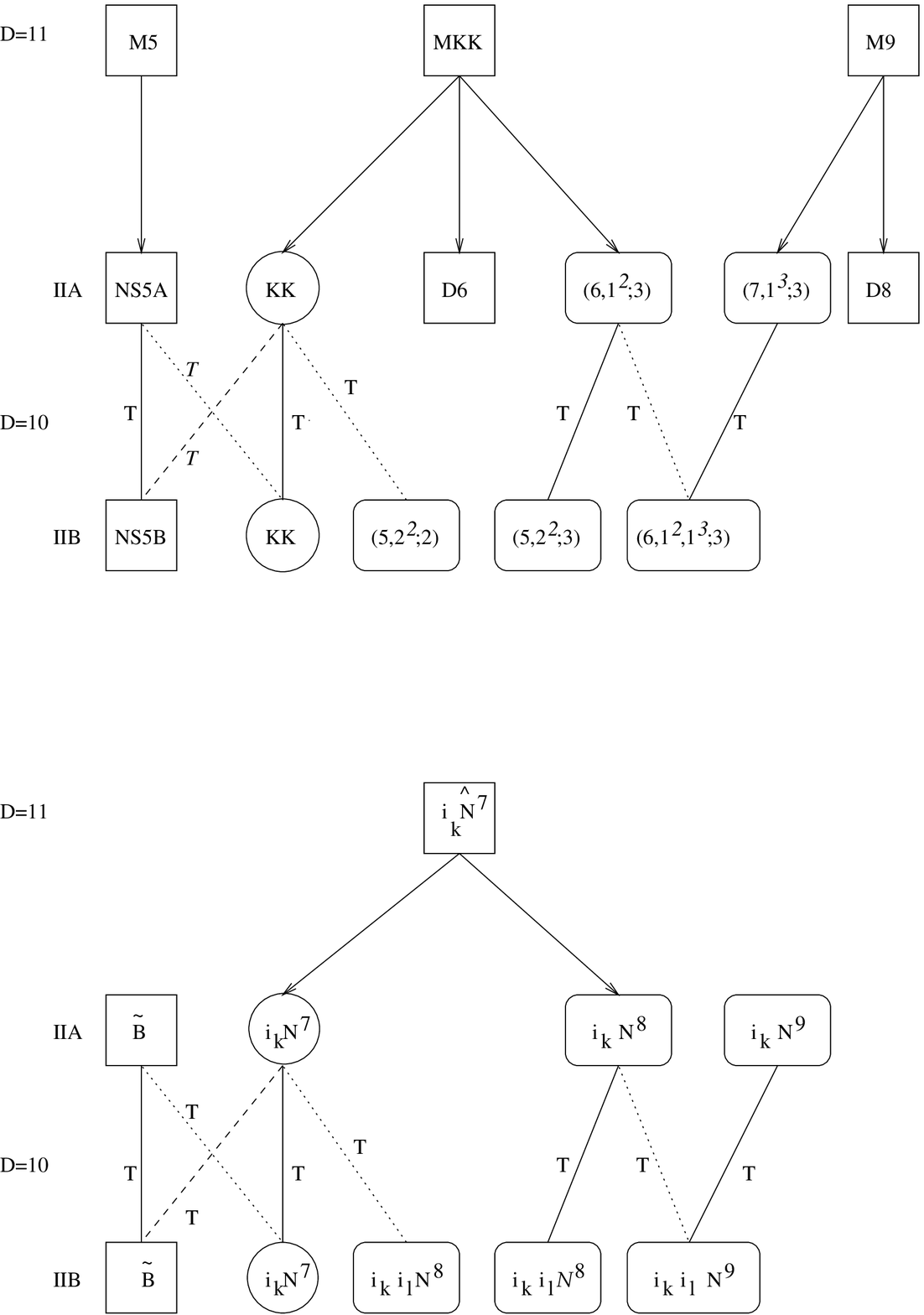}
\caption{\footnotesize
In these figures we show the relations under T-duality between the different
gravitational branes and fields that we have encountered. We also
include the M-theory origin of the Type IIA branes. With respect to the
Type IIA branes, we indicate a T-duality along a world volume direction
by a solid line, a T-duality along a Killing direction with a dashed line,
and a T-duality along a transverse direction with a dotted line.}
\label{exotics}
\end{center}
\end{figure}

We have summarised in Figure \ref{exotics} the duality relations 
between the different exotic branes,
as well as between exotic and ordinary\footnote{Predicted by the SUSY 
algebras.} branes, and we have also included 
their M-theory origins. The duality relations among the background
fields that couple minimally to these branes are also indicated in 
this figure. These connections are of particular relevance for the
calculations that we have carried out in the paper.

\begin{table}[!ht]
\begin{center}
\begin{tabular}{|c|c|}
\hline
Background field  & Brane    \\ 
\hline\hline 
${\tilde B}$ & NS5 \\ 
\hline
$i_k N^{(7)}$ &~ KK-monopole ~ \\ 
\hline
$i_k N^{(8)}$ & $(6,1^2;3)$ \\ 
\hline
$i_k N^{(9)}$ & $(7,1^3;3)$   \\
\hline
\end{tabular}
\end{center}  
\caption{\label{table1} {\footnotesize Higher form NS-NS and
gravitational background fields of the Type IIA 
superstring}}
\end{table}
In Table \ref{table1}, we summarise the higher form NS-NS and gravitational
background fields of the Type IIA superstring, and the branes, some of them
exotic, to which they couple minimally. Our notation for the exotic
branes is that of \cite{Hull}. The first entry gives the number of
ordinary spatial world volume directions, the entries of the form $m^n$ 
indicate that there are $m$ spatial directions which are gauged in the effective
action and whose radii have $n$th power. Finally the last entry gives the
power of the inverse of the string coupling constant. In this notation the
KK-monopole is, for instance, denoted as $(5,1^2;2)$. $N^{(7)}$ couples in
the world volume of the KK-monopole contracted with the Killing vector $k^\mu$,
 associated to its, isometric, Taub-NUT direction.
Similarly, $N^{(8)}$ and $N^{(9)}$ couple to 
exotic branes, contracted with the Killing vector associated
to the isometry of the brane. $N^{(7)}$ is Poincar\'e dual to the
Killing vector considered as a 1-form, $k_\mu$ \cite{BEL}, $N^{(8)}$ to
its modulus and $N^{(9)}$ to the mass of the massive Type IIA supergravity
that is obtained by reducing the massive eleven-dimensional supergravity
of \cite{BLO} along a direction different from the Killing direction
\cite{EL}.
\begin{table}[!ht]
\begin{center}
\begin{tabular}{|c|c|}
\hline
Background  field  & Brane    \\
\hline\hline
${\tilde B}$ & NS5 \\
\hline
$i_k N^{(7)}$ & ~ KK-monopole ~ \\
\hline
$i_k i_l N^{(8)}$ & $(5,2^2;3)$ \\
\hline
$i_k i_l {\cal N}^{(8)}$ & $(5,2^2;2)$ \\
\hline
$i_k i_l N^{(9)}$ & $(6,1^2,1^3;3)$   \\
\hline
\end{tabular}
\end{center}  
\caption{\label{table2} {\footnotesize Higher form NS-NS and
gravitational background fields of the Type IIB 
superstring.}}
\end{table}

In Table \ref{table2} we summarise some of the higher form NS-NS and 
gravitational
background fields of the Type IIB superstring and the branes, some of them
exotic, to which they couple minimally. The $(5,2^2;3)$-brane is
related by T-duality to the Type IIA $(6,1^2;3)$-brane, with the T-duality
taking place along a world volume direction of the IIA brane, 
direction which becomes
isometric in the dual brane. The Type IIB $(5,2^2;2)$-brane on the other hand
is related by T-duality to the Type IIA Kaluza-Klein monopole, with T-duality
taking place along a transverse direction to the monopole. This direction
becomes also isometric in the dual brane. Finally, the Type IIB 
$(6,1^2,1^3;3)$-brane is related via T-duality along a transverse direction
to the $(7,1^3;3)$-brane of Type IIA, with the duality direction
becoming again isometric. In all these cases the $N^{(8)}$, ${\cal N}^{(8)}$
and $N^{(9)}$ fields couple to the world volume of the branes contracted with the
two Killing vectors $k^\mu$ and $l^\mu$, associated to the two isometries 
of the brane. $N^{(8)}$ and ${\cal N}^{(8)}$ are Poincar\'e duals to the
two scalars $k^2$, $l^2$. The field $N^{(9)}$ is dual to a mass parameter,
which should be the one of the nine-dimensional Type II massive
supergravity with a Killing isometry to which this 7-brane gives mass
(see \cite{EL}).

\vskip 12pt


\end{document}